\documentclass[showpacs,preprintnumbers,superscriptaddress,amsmath,amssymb,pre]{revtex4}
\usepackage{graphicx}% Include figure files
\usepackage{dcolumn}% Align table columns on decimal point
\usepackage{bm}% bold math
\usepackage{color,multirow}% color
\usepackage{url}

\begin{document}
%\begin{CJK*}{GBK}{Song} % Use default fonts from CJK (see below)

%\preprint{submitted to Physical Review E}

\title{The US stock market leads the Federal funds rate and Treasury bond yields}

\author{Kun Guo}
 \affiliation{Research Center on Fictitious Economics and Data Science, Chinese Academy of Sciences, Beijing 100190, China}%
\author{Wei-Xing Zhou}
 \email{wxzhou@ecust.edu.cn}
 \affiliation{Research Center on Fictitious Economics and Data Science, Chinese Academy of Sciences, Beijing 100190, China}%
 \affiliation{School of Business, East China University of Science and Technology, Shanghai 200237, China} %
 \affiliation{School of Science, East China University of Science and Technology, Shanghai 200237, China} %
 \affiliation{Research Center for Econophysics, East China University of Science and Technology, Shanghai 200237, China} %
\author{Si-Wei Cheng}
 \affiliation{Research Center on Fictitious Economics and Data Science, Chinese Academy of Sciences, Beijing 100190, China}%
\author{Didier Sornette}
 \email{dsornette@ethz.ch}
 \affiliation{Department of Management, Technology and Economics, ETH Zurich, Kreuzplatz 5, CH-8032 Zurich, Switzerland} %
 \affiliation{Swiss Finance Institute, c/o University of Geneva, 40 blvd. Du Pont d'Arve, CH 1211 Geneva 4, Switzerland}%

\date{\today}

\begin{abstract}
Using a recently introduced method to quantify the time varying lead-lag dependencies
between pairs of economic time series (the thermal optimal path method),
we test two fundamental tenets of the theory of fixed income: (i)
the stock market variations and the yield changes should be anti-correlated;
(ii) the change in central bank rates, as a proxy of the monetary policy of the central bank,
 should be a predictor of the future stock market direction.
 Using both monthly and weekly data, we found very similar lead-lag dependence between the S\&P500 stock market index
 and the yields of bonds inside two groups: bond yields of short-term
 maturities (Federal funds rate (FFR), 3M, 6M, 1Y, 2Y, and 3Y) and bond yields of
 long-term maturities (5Y, 7Y, 10Y, and 20Y). In all cases, we
observe the opposite of (i) and (ii). First, the stock market
and yields move in the same direction.
Second, the stock market leads the yields, including and especially the FFR.
Moreover, we find that the short-term yields in the first group
lead the long-term yields in the second group before the financial crisis
that started mid-2007 and the inverse relationship holds afterwards.
These results suggest that the Federal Reserve is increasingly mindful
of the stock market behavior, seen at key to the recovery and health of the economy.
Long-term investors seem also to have been more reactive
and mindful of the signals provided by the financial stock markets than the Federal Reserve itself after the start of the
financial crisis. The lead of the  S\&P500 stock market index over
the bond yields of all maturities is confirmed by the traditional lagged cross-correlation analysis.
\end{abstract}


\maketitle

%\end{CJK*}

\section{Introduction}
\label{S1:intro}

Financial markets play a more and more important role in the economic system. Many financial variables have predictive power for output or inflation of the real economy.
Financial markets are becoming increasingly important to the real economy
due to their impact on output growth and inflation, among others
\cite{Fama-1981-AER,Arestis-Demetriades-Luintel-2001-JMCB,Estrella-Mishkim-1998-RES,Forni-Hallin-Lippi-Reichlin-2003-JMonE,Alfaro-Chanda-Kalemli-Sayek-2004-JIE,Gilchrist-Yankov-Zakrajsek-2009-JMonE}.
As an important part of financial markets, stock markets can be considered as economy barometers \cite{Schwert-1990-JF,Pan-2007-IJTAF}.
As a consequence, monetary policy, which is usually based on inflation target and sometimes
unemployment goals, is not independent of stock markets.
There is a large financial economic literature concerned with the impact of and relationship
between the monetary policy of central banks and the performance of stock markets.
The  common wisdom asserts that (i) the stock market variations and
bond yield changes should be anti-correlated and (ii)
the change in short-term interest rates, as a proxy of the monetary policy of the central bank,
 should be a predictor of the future stock market direction. The first assertion reflects
 the impact of capital cost on economic growth. The second statement is a corollary
 of the causal effect of the former one.

Some of the most relevant results for our study that were obtained by previous scholars on these two statements include the following.
Tobin's portfolio selection theory \cite{Tobin-1969-JMCB} explained the stock price increases
observed in times when the interest rate goes down as due
to investors' preference for the higher yield of stock markets. Rigobon and Sack \cite{Rigobon-Sack-2004-JMonE} documented
that an increase in short-term interest rate results in a decline in stock prices and in an upward shift
and flatter yield curve. Bernanke and Kuttner \cite{Bernanke-Kuttner-2005-JF} found that a hypothetical
unanticipated 25-basis-point cut in the FFR target is associated with approximately a 1\% increase in the broad stock indexes.
Bj{\o}rnland and Laitemo \cite{Bjornland-Leitemo-2009-JMonE} found a significant relationship, which is however
the inverse of (i) and (ii): a one percent increase of the stock market leads on average to a
4-basis-point increase of the interest rate. Two of us have also previously found that the stock market
seems to influence the FFR, during the 2000-2003 US stock market antibubble \cite{Zhou-Sornette-2004b-PA,Sornette-Zhou-2005-QF,Zhou-Sornette-2006-JMe,Zhou-Sornette-2007-PA}.

Here, using an extension of the so-called TOP technique \cite{Zhou-Sornette-2004b-PA,Sornette-Zhou-2005-QF,Zhou-Sornette-2006-JMe,Zhou-Sornette-2007-PA} for the joint analysis of pairs of time series, we revisit the pertinence of these two assertions (i) and (ii) by estimating the
lead-lag structure between the US stock market proxied by the S\&P 500 index and a set of
Treasury bond yields, including the
Federal funds rate (FFR), which constitutes one of the tools implementing monetary policy in the US.
Our analysis is applied to monthly and weekly data  of Federal funds effective rate (FFR), and nine Treasury bond yields with different maturities: 3M (3 months), 6M, 1Y (1 year), 2Y, 3Y, 5Y, 7Y, 10Y, and 20Y.
The period of analysis from August 2000 to February 2010 includes the bearish market up to mid-2003, the bullish bubble-like market
regime up to October 2007 followed by the turbulent phases associated with the so-called great Recession \cite{Reinhart-Rogoff-2009-AER,Dooley-Hutchison-2009-JIMF}. Given the extraordinary developments associated with the financial crises followed by economic crises
in different parts of the world, it is particularly interesting to investigate the
 lead-lag structure between the US stock market proxied by the S\&P 500 index and a set of
Treasury bond yields.

This work is organized as follows. In Section \ref{S1:TOPmethod}, we brief review the thermal optimal path method and propose to assess the statistical significance of the extracted lead-lag structure by using bootstrapping techniques. Section \ref{S1:Data} describes the data sets and presents the results of unit root tests for these time series.  Our main results are presented in
Section \ref{S1:Result}. Comparisons with the traditional lagged cross-correlation analysis are given in Section \ref{S1:Xcorr}. Section \ref{S1:Conclusion} summarizes our findings.

\section{Methodology}
\label{S1:TOPmethod}

\subsection{Description of the thermal optimal path (TOP) method}

The  thermal optimal path (TOP) method has been proposed as a new method to
identify and quantify the time-varying lead-lag structure between two time series.
The TOP method was successfully applied to several economic cases \cite{Sornette-Zhou-2005-QF,Zhou-Sornette-2006-JMe,Zhou-Sornette-2007-PA}. It works as follows.

Consider two standardized time series $\{X(t_1):t_1=1,...,N\}$ and $\{Y(t_2):t_2=1,...N\}$.
The matrix $E_{X,Y}$ of distances between $X$ to $Y$ is defined
as \cite{Sornette-Zhou-2005-QF,Zhou-Sornette-2006-JMe}
\begin{equation}
  \epsilon(t_1,t_2) = [X(t_1)-Y(t_2)]^2~.
  \label{Eq:DistMatrix}
\end{equation}
The element $\epsilon(t_1,t_2)$ of the matrix $E_{X,Y}$ thus compares the realization $X(t_1)$
of $X$ at time $t_1$ with the realization $Y(t_2)$ of $Y$ at time $t_2$.
The value $[X(t_1)-Y(t_2)]^2$ defines the distance between the realizations of the first time series at time $t_1$ and the second time series at time $t_2$. The $N\times N$ matrix $E_{X,Y}$ thus embodies
all possible point-wise pairwise comparisons between the two time series.
Note that the distance matrix could be modified to deal with two non-monotonic time series, for which the TOP algorithm is essentially the same \cite{Zhou-Sornette-2007-PA}.

\begin{figure}[htb]
\centering
  \includegraphics[width=8cm]{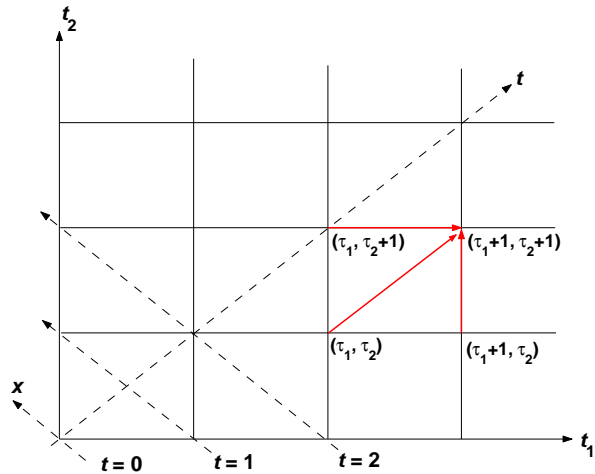}
  \caption{(Color online) Representation of the two-layer approach in the lattice $(t_1,t_2)$ and of the rotated frame $(t,x)$ as defined in the text. The three arrows depict the three moves that are allowed to reach any node in one step. }
  \label{Fig:TOP:TMM}
\end{figure}

Once the matrix $E_{X,Y}$  with elements given by Eq.~(\ref{Eq:DistMatrix}) is obtained,
an optimal path is determined that quantifies the  lead-lag dependence between the two time series.
Figure \ref{Fig:TOP:TMM} gives a schematic representation of how lead-lag paths are defined \cite{Sornette-Zhou-2005-QF}. The first (resp. second) time series is indexed by the time $t_1$ (resp. $t_2$). The nodes of the plane carry the values of the distance for each pair $(t_1,t_2)$. The path along the diagonal corresponds to taking $t_1=t_2$, i.e., compares the two time series at the same time. Paths above (resp. below) the diagonal correspond to the second time series lagging behind (resp. leading) the first time series. The figure shows three arrows which define the three causal steps (time flows from the past to the future both for $t_1$ and $t_2$) allowed in our construction of the lead-lag paths. A given path selects a contiguous set of nodes from the lower left to the upper right. The relevance or quality of a given path with respect to the detection of the lead-lag relationship between the two time series is quantified by the sum of the distances along its length, called the ``cost'' of the path.  The lead-lag structure is then obtained as the relationship $t_2(t_1)$ as a function of $t_1$, as described shortly. We stress that the two-layer scheme presented in Fig.~\ref{Fig:TOP:TMM} performs better than multi-layer schemes \cite{Zhou-Sornette-2006-JMe}.

As shown in Fig. \ref{Fig:TOP:TMM}, it is convenient to use the rotated coordinate system $(x,t)$ such that
\begin{equation}
\left\{
   \begin{array}{ccl}
    t_1 &=& 1+\left(t-x\right)/2 \\
    t_2 &=& 1+\left(t+x\right)/2
    \end{array}
\right., \label{Eq:AxesTransform2}
\end{equation}
where $t$ is in the main diagonal direction of the $(t_1,t_2)$ system and $x$ is perpendicular to $t$. The origin $(x=0,t=0)$ corresponds to $(t_1=1,t_2=1)$. Then, the standard reference path is the diagonal of equation $x=0$, and paths which have $x(t) \neq 0$ define varying lead-lag patterns. Inverting (\ref{Eq:AxesTransform2}),
we have
\begin{equation}
x = t_2 - t_1.
\label{hrtujyrmnb}
\end{equation}
This means that a positive $x$ corresponds to $t_2 > t_1$, which by definition of the
optimal thermal path below means that the second time series $Y(t_2)$ lags
behind the first time series $X(t_1)$, or equivalently  $X(t_1)$ leads $Y(t_2)$.

The idea of the TOP method is to identify the lead-lag relationship between two time series as the best path in a certain sense. A natural idea is that
the best path is the one which has the minimum sum of its distances along its length (paths are constructed with equal lengths so as to be comparable). This path with minimum
cost has thus the minimum average distance between the two time series, i.e., it is such that $Y(t_2)$ resembles the most $X(t_1)$ along this path $t_2(t_1)$.
The problem with this idea is that the noises decorating the two time series introduce spurious patterns which may control the determination the path which minimizes the sum of distances, leading to incorrect inferred lead-lag relationships. It has been shown that a robust lead-lag path is obtained by defining an average over many paths, each weighted according to a Boltzmann-Gibbs factor, hence the name ``thermal'' optimal path method \cite{Sornette-Zhou-2005-QF,Zhou-Sornette-2006-JMe,Zhou-Sornette-2007-PA}. Intuitively, this corresponds to performing
an averaging operation over neighboring paths of almost the same cost.

Concretely, we first calculate the partition functions $G(x,t)$, for all values of $x$ at a fixed $t$
in the lattice shown in Fig. \ref{Fig:TOP:TMM},
and their sum $G(t)=\sum_x G(x,t)$ so that $G(x,t)/G(t)$ can be interpreted as the probability for a path to be at distance $x$ from the diagonal for a distance $t$ along the diagonal. This probability $G(x,t)/G(t)$ is determined as a compromise between minimizing the mismatch or cost as defined above (similar to an ``energy'') and maximizing the combinatorial weight of the number of paths with similar mismatches in a neighborhood (similar to an ``entropy''). As illustrated in Fig. \ref{Fig:TOP:TMM}, in order to arrive at $(t_1+1, t_2+1)$, a path can come from $(t_1+1, t_2)$ vertically, $(t_1, t_2+1)$ horizontally, or $(t_1, t_2)$ diagonally. The recursive equation on $G(x,t)$ is therefore
\begin{equation}\label{Eq:RecurG:xt}
      G(x,t+1) = [G(x-1,t)+ G(x+1,t)+G(x,t-1)]e^{-\epsilon(x,t)/T},
\end{equation}
where $\epsilon(x,t)$ is defined by Eq.~(\ref{Eq:DistMatrix}). The parameter $T$ plays the role of
a ``temperature'' controlling the relative importance of cost versus combinatorial entropy.
The larger $T$ is, the larger the number of paths that contribute to the partition functions.
In contrast, as $T \to 0$, only the path with minimum cost counts.
The recursion relation (\ref{Eq:RecurG:xt}) is derived following the work
of Wang et al. \cite{Wang-Havlin-Schwartz-2000-JPCB}. To get $G(x,t)$ at the $t$-th layer,
we need to know and bookkeep the previous two layers from $G(\cdot,t-2)$ to $G(\cdot,t-1)$. After $G(\cdot,t)$ is determined, these values are normalized by
$G(t)$ so that $G(x,t)$ does not diverge at large $t$. The boundary condition of $G(x,t)$ plays an crucial role. For $t=0$ and $t=1$, $G(x,t) = 1$. For $t>1$, the boundary condition is taken to be $G(x=\pm t,t) = 0$, in order to prevent paths to remain on the boundaries.

Once the partition functions $G(x,t)$'s have been calculated, we can obtain any statistical average related to the positions of the paths weighted by the set of $G(x,t)$'s. For instance, the local time lag $\langle{x(t)}\rangle$ at time $t$ is given by
\begin{equation}
    \langle{x(t)}\rangle = \sum_x {xG(x,t)/G(t)}~.
    \label{Eq:Xave}
\end{equation}
Expression (\ref{Eq:Xave}) defines $\langle{x(t)}\rangle$ as the thermal average of the local time lag at $t$ over all possible lead-lag configurations suitably weighted according to the exponential of minus the measure $\epsilon(x,t)$ of the similarities of two time series. For a given $x_0$ and temperature $T$, we determine the thermal optimal path $\langle{x(t)}\rangle$. We can also define an ``energy'' or cost $e_T(x_0)$ to this path, defined as the thermal average of the measure $\epsilon(x,t)$ of the similarities of two time series:
\begin{equation}\label{Eq:e}
        e_T(x_0) = \frac{1}{2(N-|x_0|)-1}\sum_{t=|x_0|}^{2N-1-|x_0|}
        \sum_x {\epsilon(x,t)G(x,t)/G(t)}~.
\end{equation}

\subsection{Bootstrapping tests and statistical significance \label{tenbe}}

In order to test whether the extracted lead-lag structure is statistically significant, we introduce
a bootstrap approach \cite{Beran-1988-JASA} that is specifically adapted to the present problem.
This statistical test extends and make more robust the method and results, as compared with
previous works \cite{Sornette-Zhou-2005-QF,Zhou-Sornette-2006-JMe,Zhou-Sornette-2007-PA}.
Consider two time series $X(t_1)$ (for instance the logarithmic returns of S\&P 500) and $Y(t_2)$ (for instance the time increments of bond yields). We perform the TOP analysis on a fixed time interval at some temperature
$T$. Let us assume we obtain the lead-lag function $x(t)$. Recall that $t$ is the diagonal of the $t_1\times t_2$ plane. We then shuffle $X(t)$ and $Y(t)$ and redo the TOP analysis at the same temperature $T$. We obtain a new lead-lag function $x_1(t)$. This process is repeated another $n-1$ times, giving a total of $n$ paths $x_i(t)$ with $i=1,2,\ldots,n$. A typical value of $n$ used below is 1000.
%\begin{equation}
 %\bar{x}(t) = \frac{1}{n}\sum_{i=1}^n x_i(t).
 %\label{Eq:Bootstrap:Ave}
%\end{equation}
For each $t$, out of the $n=1000$ reshuffled time series,
we determine the 5\% quantile $x_{5\%}(t)$ and the 95\% quantile $x_{95\%}(t)$, denoted in the
following as $L(t)$ and $U(t)$. If $x(t)$ is smaller than $L(t)$ or larger than $U(t)$, we interpret that the lead-lag $x(t)$ at time $t$ is different from zero at the significance level of 95\% or larger. Complementarily,
given the obtained lead-lag $x(t)$, out of the $n=1000$ reshuffled time series, we obtain the $p$-value
as a function of $t$, which thus characterizes the time periods when there is a statistically significant
lead-lag structure as those with small $p$-values.

\section{Description of the data sets}
\label{S1:Data}

\subsection{Data sets}

In the following, we apply the TOP method respectively to monthly and weekly data of the S\&P 500 index, Federal funds effective rate (FFR), and nine Treasury bond yields with different maturities: 3M (3 months), 6M, 1Y (1 year), 2Y, 3Y, 5Y, 7Y, 10Y, and 20Y. Each time series spans from August 2000 to February 2010. The Treasury bond at 30-year maturity is not considered because it was discontinued in January 2002 and then reintroduced in February 2006.

\begin{figure}[htb]
 \centering
 \includegraphics[width=8cm,height=6cm]{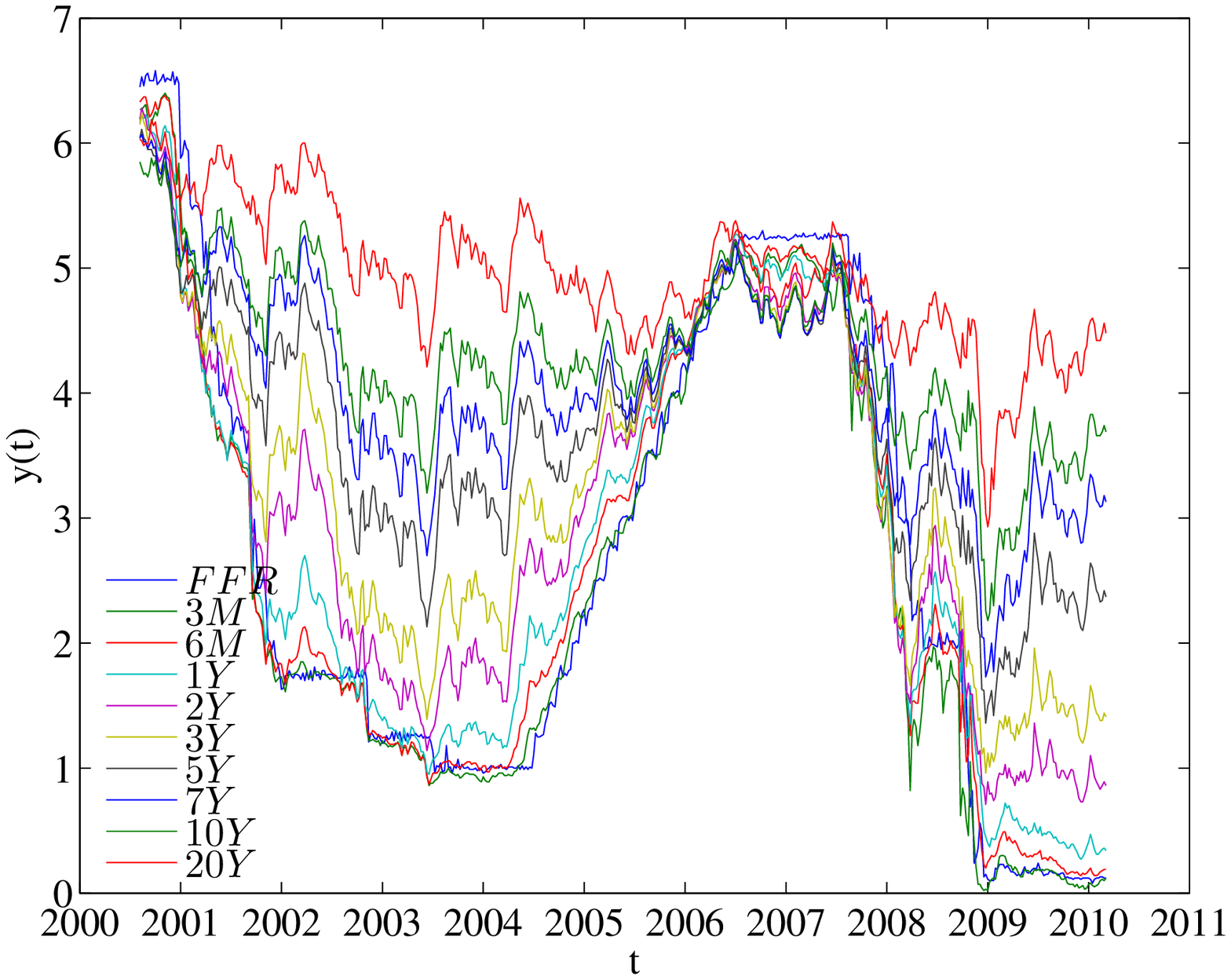}
 \includegraphics[width=8cm,height=6cm]{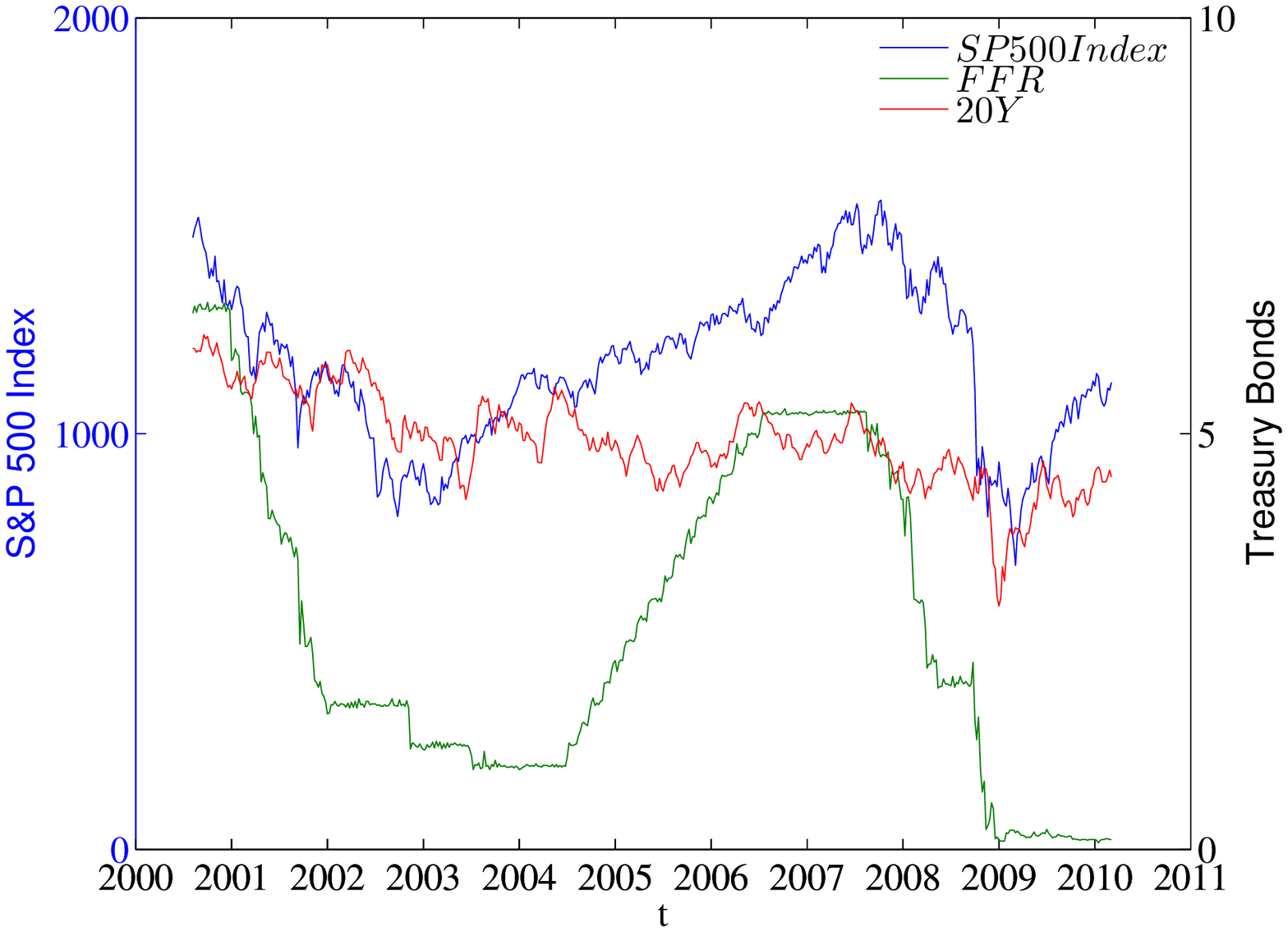}
 \caption{\label{Fig:TOP:Data} (Left) Weekly sampling of the Federal effective funds rate (FFR)
 and nine Treasury bond yields. (Right) S\&P 500 and FFR together with the 20Y for comparison.}
\end{figure}

Figure \ref{Fig:TOP:Data} shows the weekly sampling of the FFR, the nine Treasury bond yields with different maturities, and the S\&P 500 index. In the left panel of Fig.~\ref{Fig:TOP:Data}, very interesting patterns emerge in the term structure. In general, the yields of Treasury bonds with short maturities are more sensitive to the economic circumstance and change to a larger extent. In 2000, 2006 and 2007, the spread is very narrow and the FFR is even higher than the Treasury bond yields, corresponding to an anomalous inverted yield curve. These time periods correspond
respectively to the early stages of the 2000 US stock market crash and to the current financial crisis. The spread reaches local maxima in 2004 and 2010. In addition, the right panel of Fig.~\ref{Fig:TOP:Data} suggests that the FFR and the S\&P 500 index change roughly in the same direction. It is thus interesting to refine this visual impression and determine rigorously using the TOP method described above what is the lead-lag structure between the evolution of the US stock market and the FFR, which embodies an important part of the policy of the Federal Reserve.

In this paper, we use as inputs the logarithmic returns of the S\&P 500 index and the increments of the FFR and of all the yields, rather than the non-stationary original time series. We define the logarithmic returns of the S\&P 500 index as follows
\begin{equation}
 X(t) = \ln[S\&P(t)]-\ln[S\&P(t-1)],~
 \label{Eq:Xt}
\end{equation}
and the logarithmic increments of yields curves as follows
\begin{equation}
 Y(t) = \ln[{\rm yield}(t)]-\ln[{\rm yield}(t-1)]~,
 \label{Eq:Yt}
\end{equation}
where the time unit for $t$ is one week for weekly data and one month for monthly data. We then normalize the two time series $X(t)$ and $Y(t)$ so that their mean is zero and their standard deviation is equal to $1$ \cite{Sornette-Zhou-2005-QF}. This ensures that they are comparable and can be used meaningfully in the TOP analysis to extract their lead-lag structure.

\subsection{Unit root tests}

We perform unit root tests on the logarithms of the original time series and their first-order differences ($X(t)$ and $Y(t)$) to check for their stationary. The augmented Dickey-Fuller (ADF) \cite{Dickey-Fuller-1981-Em}, Phillips-Perron (PP) \cite{Phillips-Perron-1988-Bm}, and Kwiatkowski-Phillips-Schmidt-Shin (KPSS) \cite{Kwiatkowski-Phillips-Schmidt-Shin-1992-JEm} tests are adopted. For the ADF and PP tests, the null hypothesis is that the time series has a unit root, which utilizes the $t$-statistic. In contrast, the null hypothesis of the KPSS method is that the time series is stationary and uses the LM-statistic. The results are presented in Table \ref{TB:UnitRoot}.

\begin{table}[htbp]
\centering
    \caption{Unit root tests of the logarithmic monthly and weekly data and their first-order differences. The augmented Dickey-Fuller (ADF), Phillips-Perron (PP) and Kwiatkowski-Phillips-Schmidt-Shin (KPSS) tests are adopted. $V_{x\%}$ is the critical value at the $x\%$ significance level, $V$ is the statistic, and $p$ is the $p$-value.}
    \medskip
    \label{TB:UnitRoot}
    \begin{tabular}{ccccccccccccccccccccccccc}
    \hline\hline
      && \multicolumn{3}{@{\extracolsep\fill}c}{$V_{x\%}$} && \multicolumn{2}{@{\extracolsep\fill}c}{S\&P 500} && \multicolumn{2}{@{\extracolsep\fill}c}{FFR}  && \multicolumn{2}{@{\extracolsep\fill}c}{1Y} && \multicolumn{2}{@{\extracolsep\fill}c}{5Y} && \multicolumn{2}{@{\extracolsep\fill}c}{20Y}\\
   \cline{3-5} \cline{7-8} \cline{10-11} \cline{13-14} \cline{16-17} \cline{19-20}
    method&& $V_{10\%}$ & $V_{5\%}$ & $V_{1\%}$ &&  $V$  &  $p$ &&  $V$  &  $p$ && $V$ & $p$ && $V$ & $p$ && $V$ & $p$\\\hline
    &&\multicolumn{10}{@{\extracolsep\fill}l}{Logarithmic monthly data}\\
    \hline
    ADF    &&  -2.58&-2.89&-3.49&& -2.14 & 0.23 && -0.56 & 0.87 && -0.64 & 0.86 && -2.27 & 0.16 && -2.51 & 0.12\\%
    PP     &&  -2.58&-2.89&-3.49&&-2.15&0.23&&-0.01&0.95&&-0.23&0.93&&-2.00&0.27&&-2.51&0.12\\%
    KPSS   &&   0.35& 0.46& 0.74&&0.16&$\gg$0.1&&0.31&$\gg$0.1&&0.27&$\gg$0.1&&0.43&0.06&&1.02&0.00  \\%
    \hline
    &&\multicolumn{10}{@{\extracolsep\fill}l}{Logarithmic weekly data}\\
    \hline
    ADF    &&  -2.57&-2.86&-3.44 && -2.08&0.25     && -0.16&0.94 && -0.29&0.92 && -2.10&0.21 && -2.95&0.04\\%
    PP     &&  -2.57&-2.86&-3.44 && -2.08&0.25     && -0.00&0.96 && -0.30&0.92 && -2.04&0.27 && -2.72&0.07\\%
    KPSS   &&   0.35& 0.46& 0.74 &&  0.31&$\gg$0.1 &&  0.69&0.01 &&  0.62&0.02 &&  0.84&0.00 &&  1.91&0.00\\%
    \hline
    &&\multicolumn{10}{@{\extracolsep\fill}l}{Difference of logarithmic monthly data}\\
    \hline
    ADF    &&  -2.58&-2.89&-3.49 && -8.29&0.00     && -4.21&0.00     && -6.68&0.00    &&-8.07&0.00     &&-9.35&0.00\\%
    PP     &&  -2.58&-2.89&-3.49 && -8.35&0.00     && -5.36&0.00     && -6.66&0.00    &&-8.07&0.00     &&-9.83&0.00\\%
    KPSS   &&   0.35& 0.46& 0.74 &&  0.11&$\gg$0.1 &&  0.32&$\gg$0.1 &&  0.32&$\gg$0.1&& 0.10&$\gg$0.1 && 0.05&$\gg$0.1\\%
    \hline
    &&\multicolumn{10}{@{\extracolsep\fill}l}{Difference of logarithmic weekly data}\\
    \hline
    ADF    &&  -2.57&-2.86&-3.44 && -22.5&0.00     && -24.6&0.00 && -16.6&0.00     && -19.3&0.00     && -17.4&0.00\\%
    PP     &&  -2.57&-2.86&-3.44 && -22.5&0.00     && -24.6&0.00 && -16.8&0.00     && -19.4&0.00     && -17.4&0.00\\%
    KPSS   &&   0.35& 0.46& 0.74 &&  0.13&$\gg0.1$ &&  0.35&0.10 &&  0.31&$\gg0.1$ &&  0.07&$\gg$0.1 &&  0.04&$\gg$0.1 \\%
   \hline\hline
    \end{tabular}
\end{table}

For the logarithmic monthly data and logarithmic weekly data, the ADF and PP tests show that these time series are not stationary and have a unit root since the $p$-values are greater than 10\%, except for the 20Y yield. In contrast, the KPSS test suggests that four time series are stationary since the $p$-values are much greater than 10\%.

For the differences of the logarithmic monthly data and logarithmic weekly data, the ADF and PP tests show that all time series are stationary at the 1\% significance level, and the KPSS test also confirms that these time series are stationary at the 10\% level. These results justify our use of the logarithmic returns in the TOP analysis in order to avoid
possible spurious signals in the estimated lead-lag structure that could result from
large excursions exhibited by the non-stationary time series.

\section{The S\&P500 leads all yields: evidence from the TOP method}
\label{S1:Result}

\subsection{Empirical results}

\begin{figure}[htb]
 \centering
 \includegraphics[width=8cm]{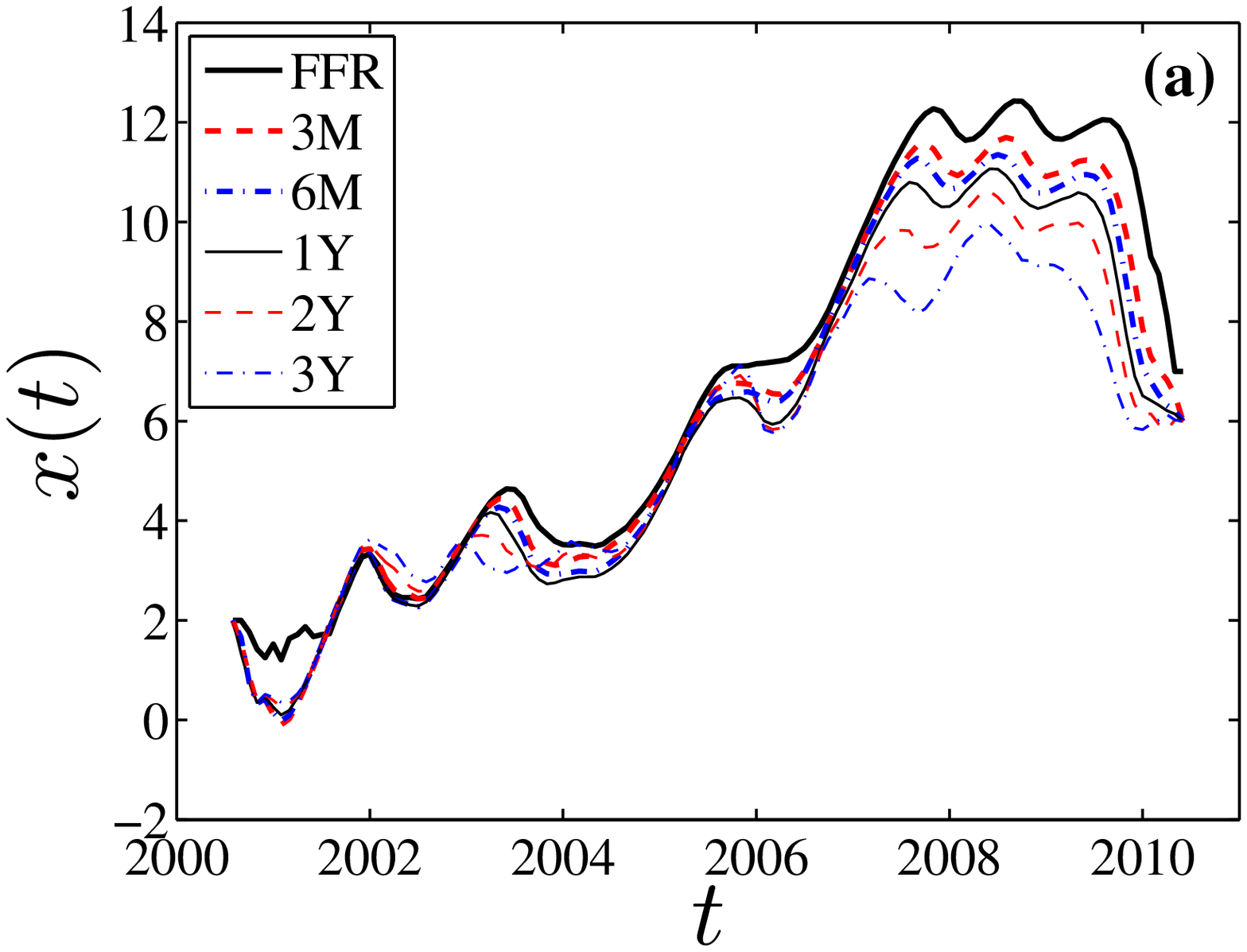}
 \includegraphics[width=8cm]{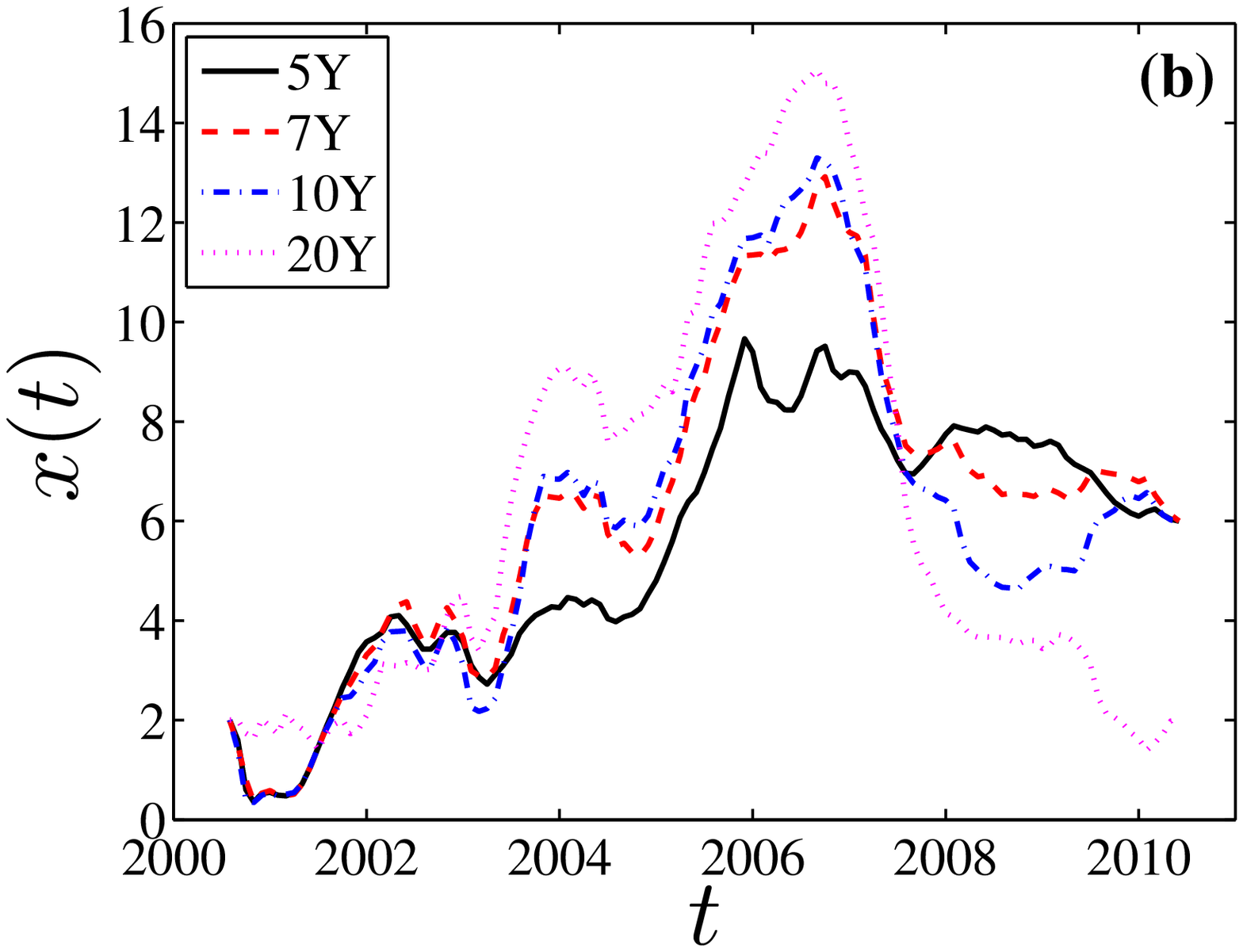}
 \caption{\label{Fig:monthly:xt} (Color online) Dependence of the lead-lag $x(t)$ between
the returns of the S\&P500 index taken as the first time series and the logarithmic variation
of each of the yields for the monthly data: (a) FFR, 3M, 6M, 1Y, 2Y, and 3Y Treasury bond yields as the first group; (b) 5Y, 7Y, 10Y, and 20Y bond yields as the second group. The unit of $x(t)$ is one month.}
\end{figure}

Figure \ref{Fig:monthly:xt} shows the instantaneous evolution of the lead-lag $x(t)$ between
the returns of the S\&P500 index taken as the first time series and the logarithmic variation
of each of the yields for the monthly data at
temperature $T=2$. We have been careful to investigate the impact of the locations of the starting
and ending extremities of the paths. There are indeed a total of $19\times19$ thermal optimal paths,
because there are 19 starting points $(t_{\rm{1,start}},t_{2,\rm{start}})$ and 19 ending points $(t_{1,\rm{end}},t_{2,\rm{end}})$, denoted using the $(t_1,t_2)$ system instead of the $(t,x)$ system for simplicity. The 19 starting points are $(t_1=0,t_2=0)$, $(t_1=0,t_2=i)$, and $(t_1=i, t_2=0)$ for $i=1,2,\ldots,9$. The 19 ending points are $(t_1=N,t_2=N)$, $(t_1=N,t_2=N-i)$ and $(t_1=N-i, t_2=N)$ for $i=1,2,\ldots,9$, where $N$ is the length of the time series. The overall thermal optimal path $x(t)$ is chosen as the one with minimal energy (or total cost) among the $19\times19$
thermal paths. As for the choice of the temperature $T$, we investigated other values and found our
results to be robust and qualitatlvely similar with respect to variations of $T$ between $1$ and $3$.
To present our results, we choose this value $T=2$ as it seems to represent a reasonable optimal,
confirmed by cross-correlation analyses performed on the steady periods
found with fixed lag times for various $T$'s.

Figure~\ref{Fig:monthly:xt} is organized in two panels, each panel plotting one group.
The first group includes FFR, 3M, 6M, 1Y, 2Y, and 3Y Treasury bonds as shown in Fig.~\ref{Fig:monthly:xt}(a). The second group includes 5Y, 7Y, 10Y, and 20Y Treasury bonds as shown in Fig.~\ref{Fig:monthly:xt}(b). The evolution of $x(t)$ in each group are quantitatively similar.

\begin{figure}[htb]
 \centering
 \includegraphics[width=8cm]{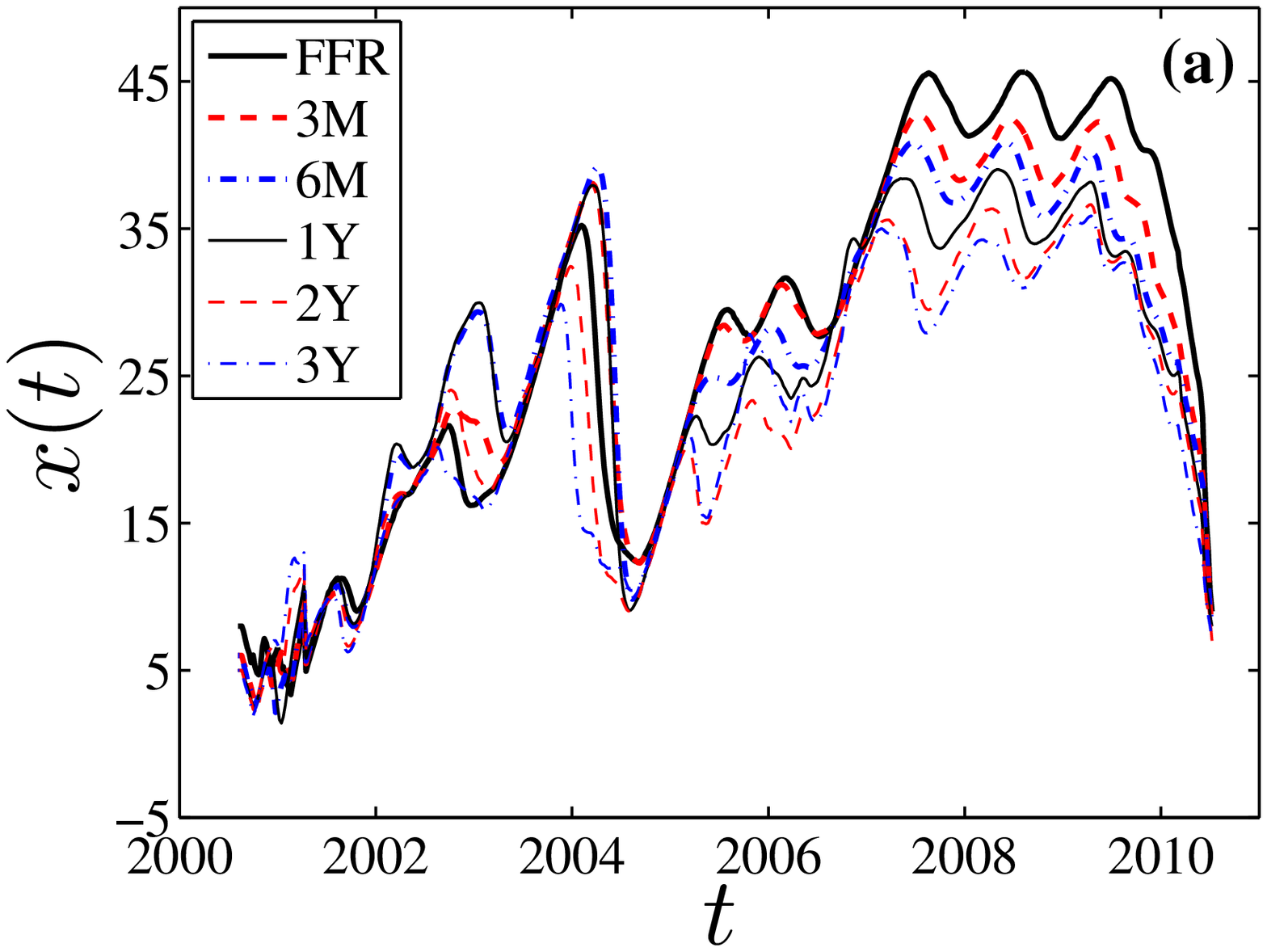}
 \includegraphics[width=8cm]{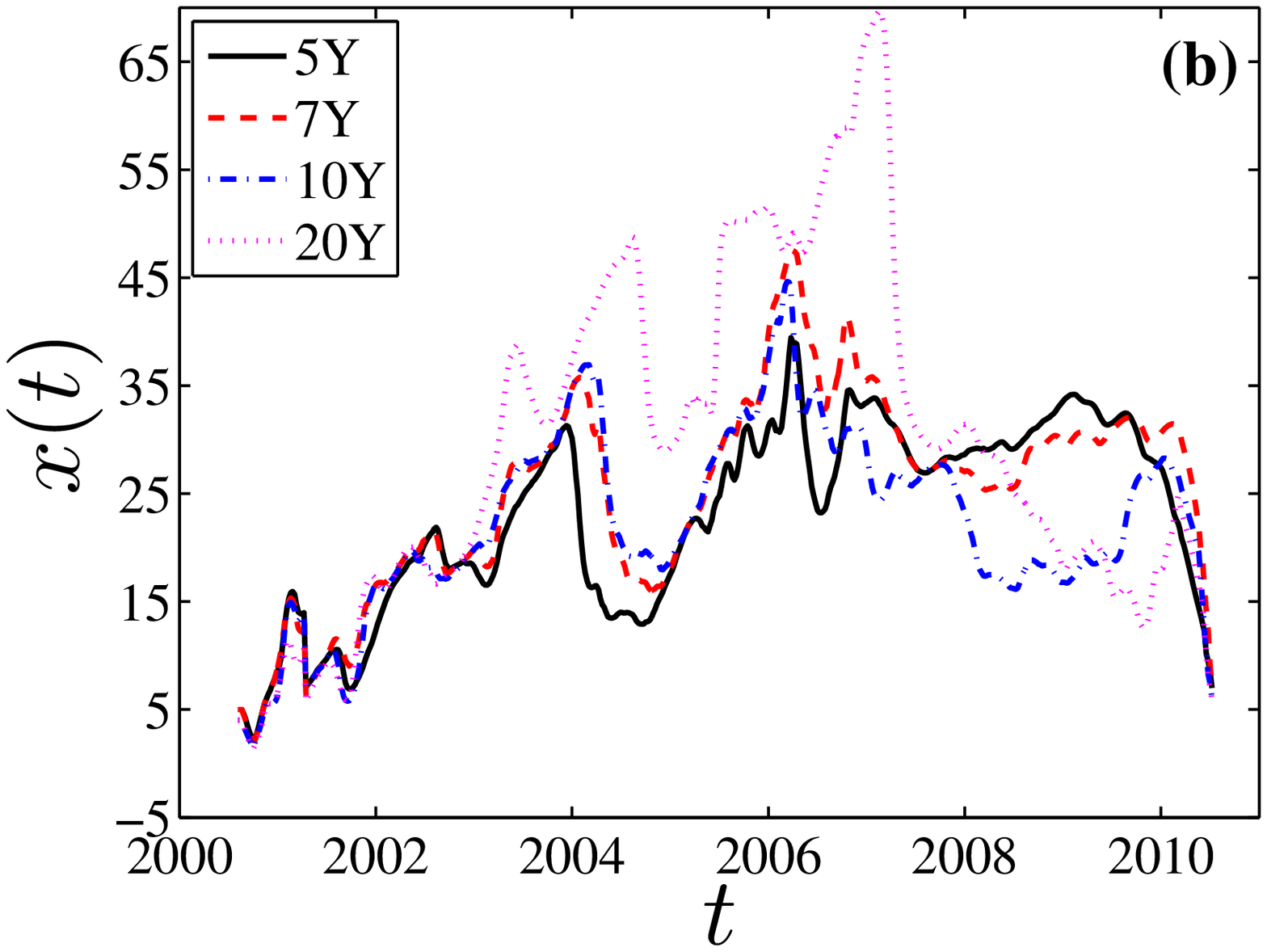}
 \caption{\label{Fig:weekly:xt} (Color online) Same as figure \protect\ref{Fig:monthly:xt} for weekly data.}
\end{figure}

Figure \ref{Fig:weekly:xt} is the same as Fig. \ref{Fig:monthly:xt} for weekly data.
Apart from largest fluctuations of the lead functions, the results are very similar and robust
to this change of time scale from monthly to weekly.

\subsection{Statistical significance}

\begin{figure}[htb]
 \centering
 \includegraphics[width=8cm]{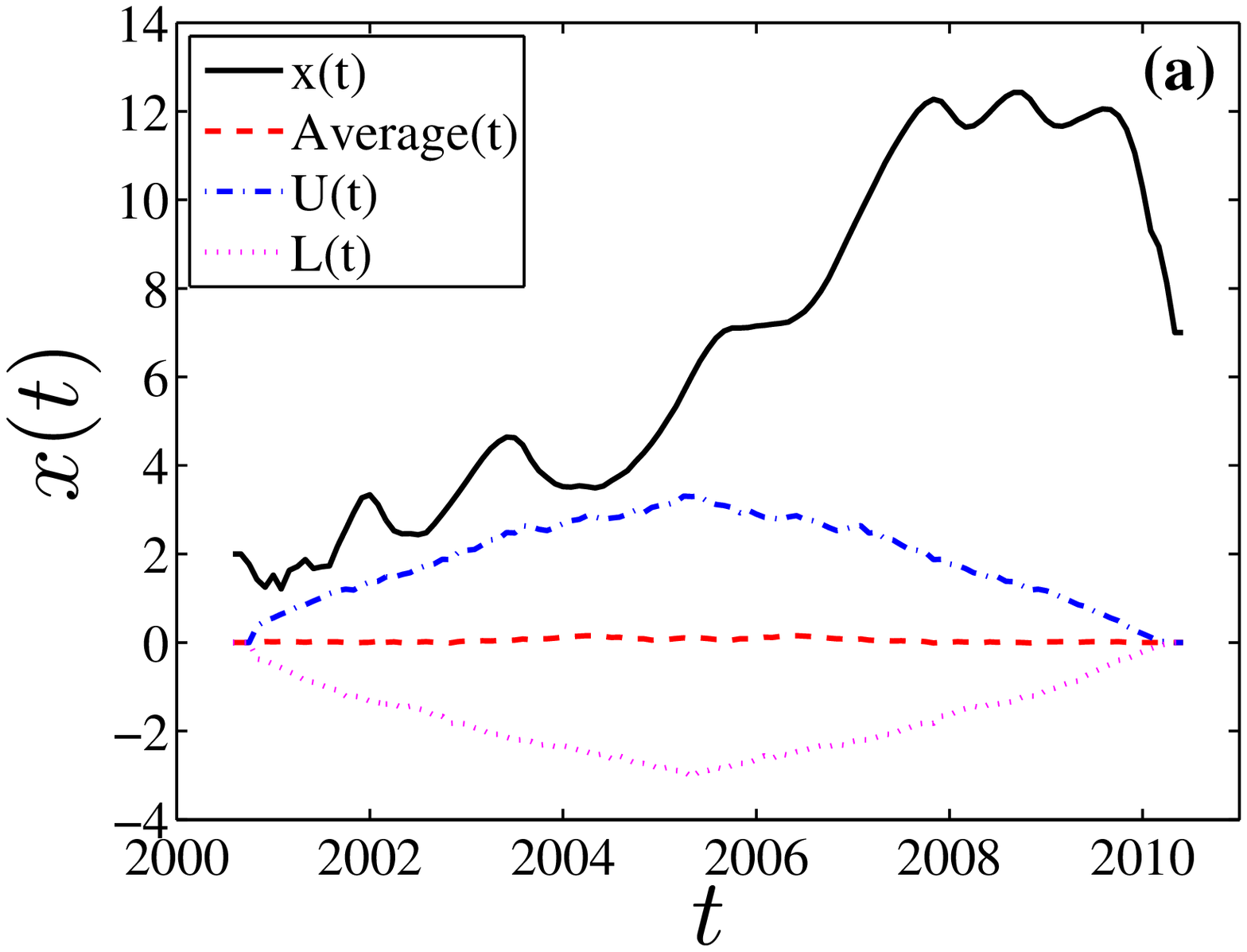}
 \includegraphics[width=8cm]{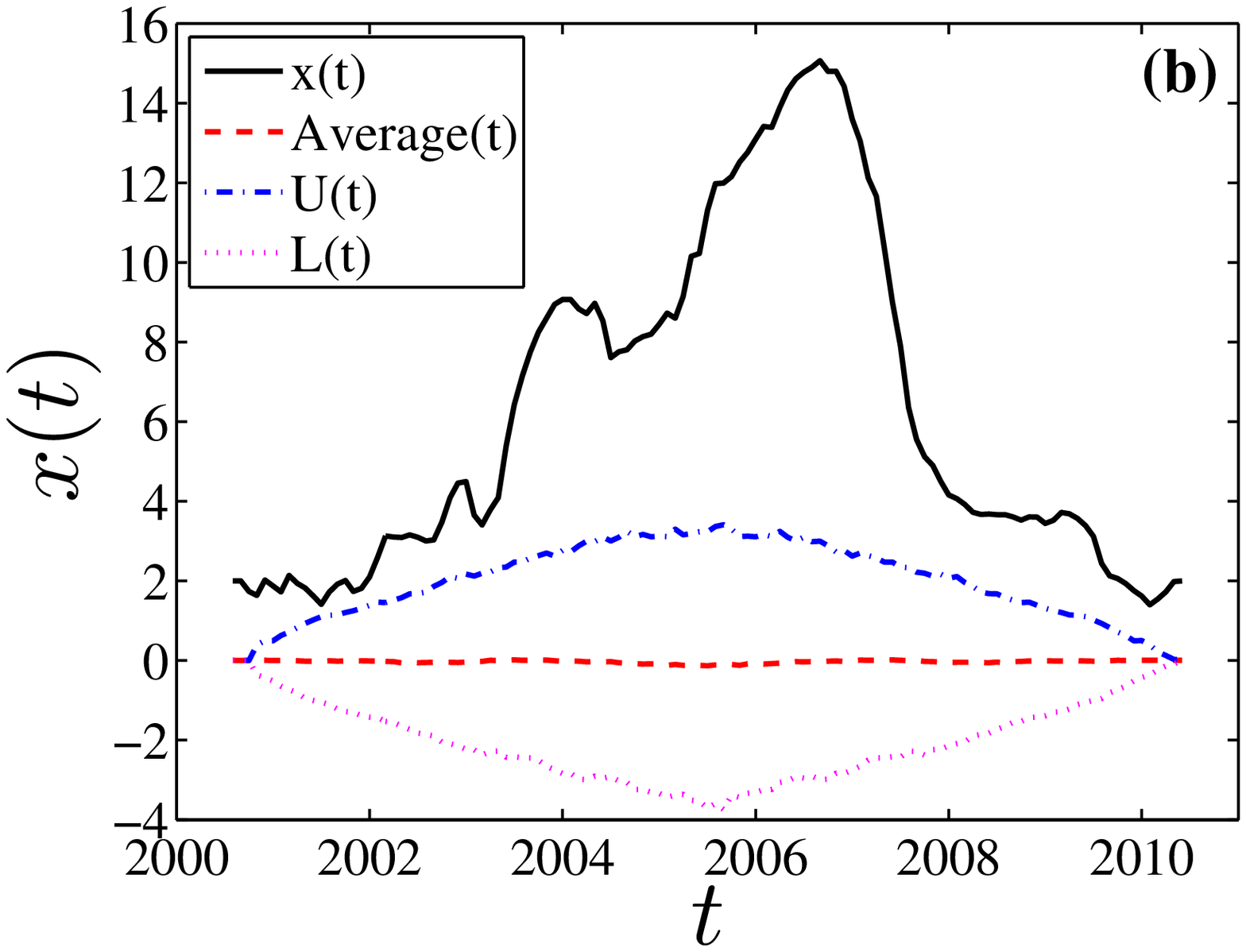}
 \caption{\label{Fig:monthly:xt:test} (Color online) Bootstrap test for the significance of the lead-lag structure for the FFR (a) and the 20Y Treasury bond yield (b).}
\end{figure}

Before commenting and exploiting the information presented in Figs.~\ref{Fig:monthly:xt} and
\ref{Fig:weekly:xt}, it is important to ascertain their statistical significance. For this, we use the bootstrap method
described above in section \ref{tenbe}.
Figure \ref{Fig:monthly:xt:test} illustrates the obtained results from the monthly data for two maturities, namely
the shortest one (FFR) and the longest one (20-year Treasury bond yield). It shows that the two
lead function $x(t)$ are well above the 95\% quantile curves, that is, $x(t)>U(t)$. The conclusion is the same for other Treasury bond yields. We conclude that the obtained lead-lag structure for the monthly data cannot be produced by chance at the 95\% significance level.

\begin{figure}[htb]
 \centering
 \includegraphics[width=8cm]{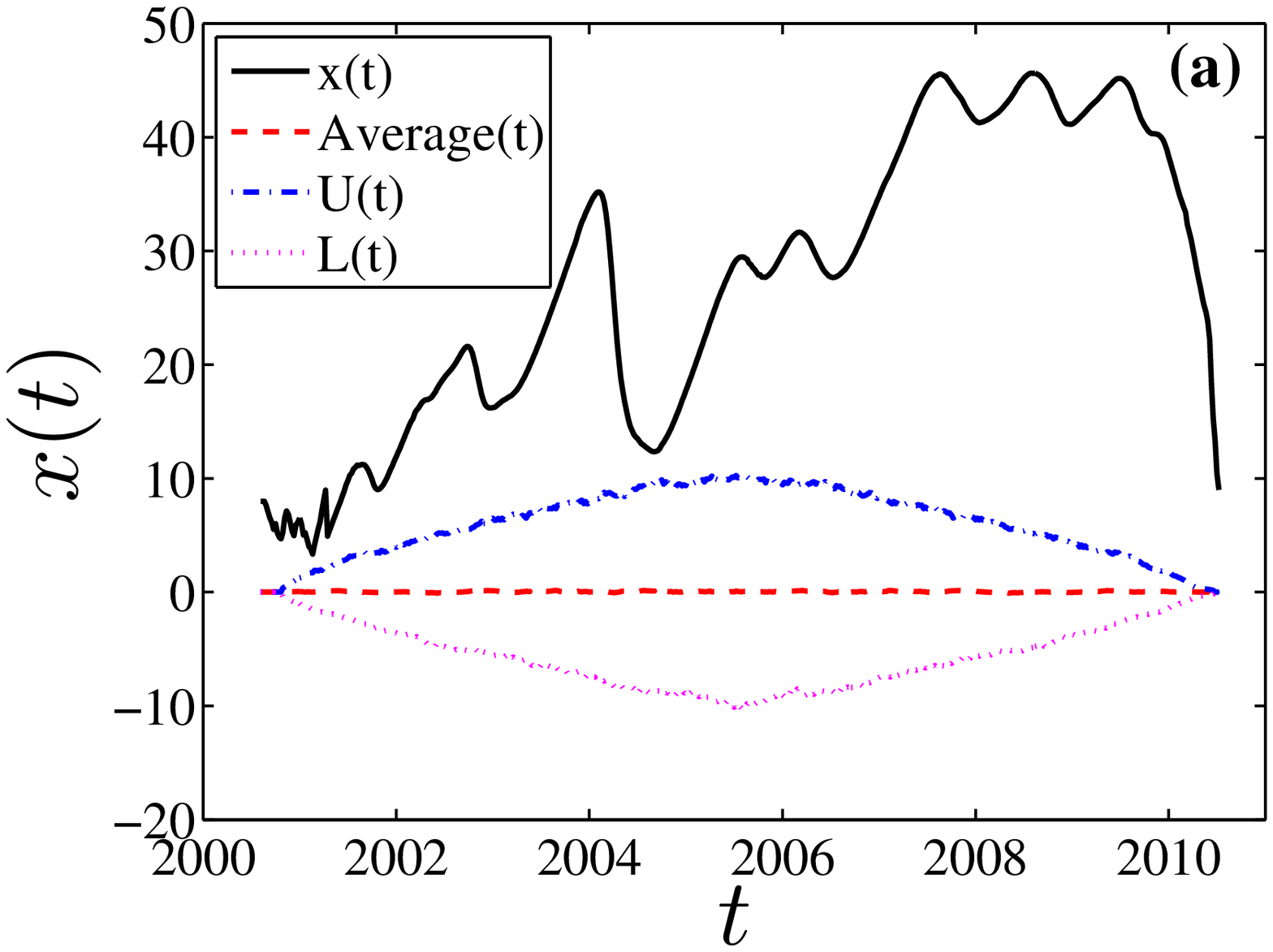}
 \includegraphics[width=8cm]{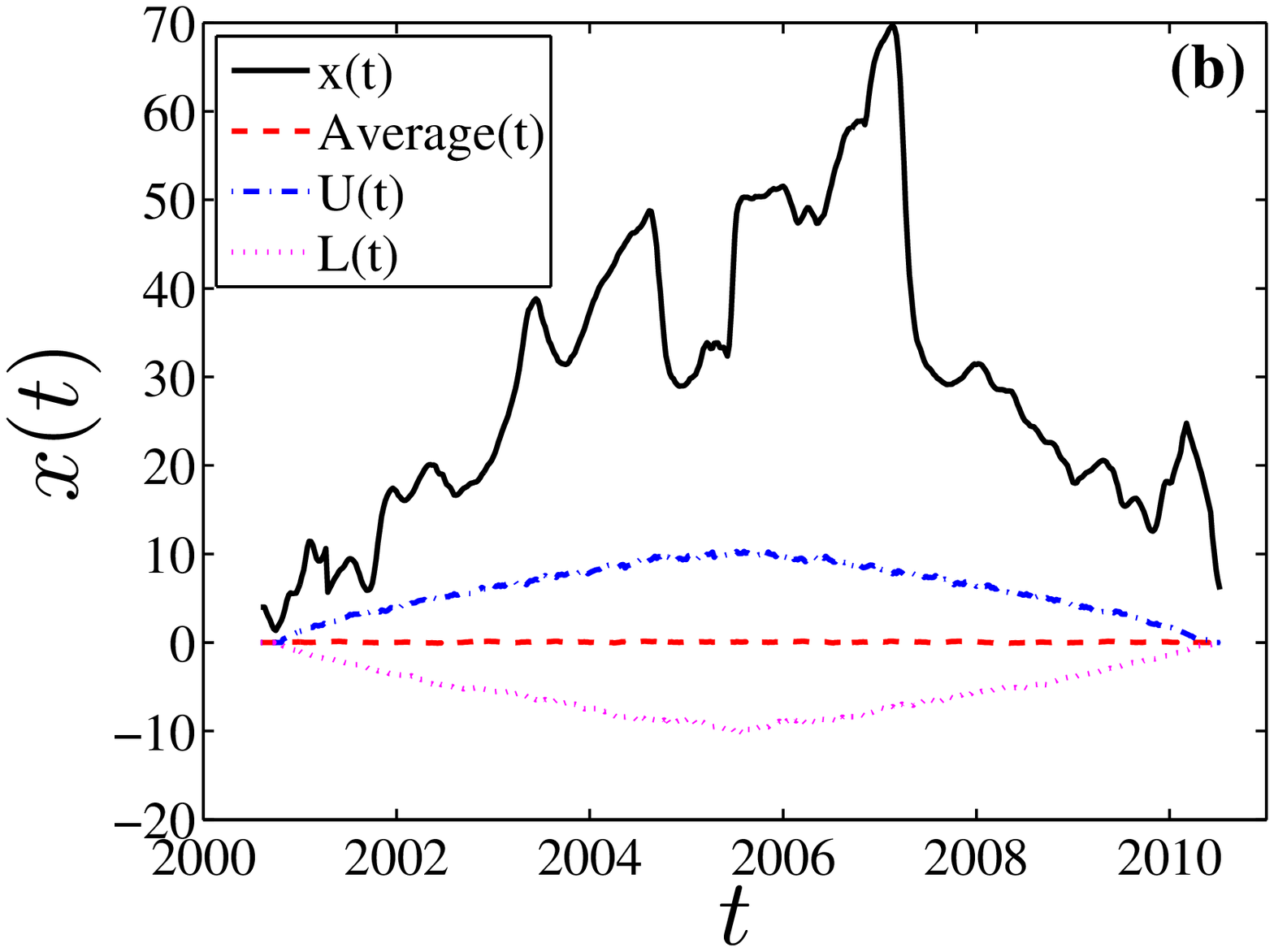}
 \caption{\label{Fig:weekly:xt:test} (Color online) Bootstrap test for the significance of the lead-lag structure for the FFR (a) and the 20Y Treasury bond yield (b).}
\end{figure}

Figure \ref{Fig:weekly:xt:test} illustrates the obtained results from the weekly data for two maturities, namely
the shortest one (FFR) and the longest one (20-year Treasury bond yield). The conclusion is the same for other Treasury bond yields. Therefore, the $x(t)$ functions for the weekly data are positive at the 95\% significance level, which unveils the nontrivial intrinsic lead-lag structure of the S\&P 500 index and the yield time series.

\subsection{Two shocking stylized facts}

The first and most important observation extracted
from  Figs  \ref{Fig:monthly:xt} and \ref{Fig:weekly:xt}
is that, for all yields and at all times, the S\&P500 index
leads the yield changes, since $x(t)$ is always positive, which by
definition (\ref{hrtujyrmnb}), means that  $t_2 > t_1$ for the optimal thermal path. Since
the index $t_1$ corresponds to the S\&P500 index and the index $t_2$ corresponds
to one of the yields, this conclusion follows. This result confirms and extends considerably that
reported previously by two of us \cite{Zhou-Sornette-2004b-PA}
using standard measures of correlations over a restricted period from
2001 to 2003, under the somewhat provocative
title ``Causal slaving of the U.S. Treasury Bond Yield ... by the
Stock Market...'' Indeed, as this title suggests, this result $x(t)>0$ is
particularly striking and rich of implication. This result collides against the
common wisdom that usually asserts the following two rules:
\begin{itemize}
\item[(i)] the stock market variations and the yield changes should be anti-correlated;
\item[(ii)] the change in FFR, as a proxy of the monetary policy of the central bank,
 should be a predictor of the future stock market direction.
\end{itemize}
Indeed, according to the standard story, a lower interest rate means lower costs of borrowing for the
private sector, implying that the private sector is going to profit from this
opportunity by increased investments in innovations and entrepreneurial opportunities,
leading (with some lag) to an improved outlook for the future growth of the economy.
Since stock market prices reflect the anticipation of investors, this better
outlook for the future economy should be soon reflected in the appreciation of the stock market.
Reciprocally, an increase of the yields and in particular
of the FFR should, according to the standard story, translate soon into
a drag on the growth of stock markets.

We observe the opposite of (i) and (ii). First, we find that the stock market
and yields move in the same direction, as pointed out independently by R. Werner \cite{Werner-2005}.
Second, the stock market leads the yields, including and especially the FFR.
The implication is clear: the central bank policy is (1) reacting to the
stock market and (2)  is following it. When the stock market exhibits a rally,
the Fed tends to progressively increase its rates as an attempt to calm down the
``overheating engine'', as occurred
towards the end of the ICT bubble when the Fed rate was increased to 6.5\%.
A similar increase of the Fed rate occurred from 2004 to 2007.
When the stock market plunges, the Fed tends to decrease its rates,
in the hope of putting a brake on the stock market losses
that negatively feedback onto the real economy via the wealth effect.

Both previous and present Fed chairmen Greenspan and Bernanke
have increasingly made clear that the Federal Reserve does
care more and more about the evolution of the stock markets.
On Dec. 3rd, 2010, former Federal Reserve Chairman Alan Greenspan told CNBC
 that rising stock values have played a critical role in the economic recovery.
 The stock market got a boost from the Fed policy to boost liquidity,
 which drove interest rates down and pushed investors toward riskier investments like stocks.
``I think we are underestimating and continuing to underestimate how important asset prices, very specifically equity prices, are not only to shareholders but the economy as a whole,'' he said.
Equities have risen more than 80\% from the lows set during the financial crisis,
noted Greenspan, benefiting investors and helping fuel the recovery.
[Source:
\url{http://www.dailyfinance.com/story/investing/greenspan-rising-stock-markets-are-key-to-recovery/19743325/?icid=sphere_copyright}].
On Nov. 3rd., 2010, Bernanke issued the following statement
in an opinion article for the Washington Post released hours after the Fed announced the
\$600 billion of Treasury buying through June in a second round of unconventional monetary stimulus:
``Resuming large-scale asset purchases should boost economic growth through lower borrowing costs and higher stock prices...
Stock prices rose and long-term interest rates fell when
investors began to anticipate this additional action... Easier financial conditions will promote economic growth.''
Being content to see
the stock market growing, this suggests a hidden mandate of the
Federal Reserve to steer the stock markets.

It seems that the dynamics of the Fed policy, as translated in
the Fed rates and the longer maturity yields (which of course are far
from being controlled by the central bank), is much more straightforward
than articulated in fancy models \cite{Baeriswyl-Cornand-2010-JMonE}.
The evidence presented here suggests that  Fed policy appears to be as if a straightforward reaction to financial markets was
the main factor.

\subsection{Comparison between different yields}

Comparing the lead functions $x(t)$ for the various yields with different maturities, we find that the short-term yields in the first group (left panel of Figs. \ref{Fig:monthly:xt} and \ref{Fig:weekly:xt}) move approximately in synchrony with the long-term yields in the second group (right panel of Figs. \ref{Fig:monthly:xt} and \ref{Fig:weekly:xt}) until 2007. And this synchrony is almost perfect from the yields spanning FFR to 3Y
in the first group until mid-2007. Thereafter, during the time period following the financial crisis that started in mid-2007, we can observe that the short-term yields clearly lead the long-term yields and we have the sequence of inequalities
\begin{equation}
  x_{\rm{FFR}}(t) > x_{\rm{3M}}(t) > x_{\rm{6M}}(t)>x_{\rm{1Y}}(t)>x_{\rm{2Y}}(t)>x_{\rm{3Y}}(t)>0~.
  \label{Eq:TOP:Month:xt:G1}
\end{equation}
This is seen from the fact that $x(t)$ tends to be larger for
the short-term yields, since they are all compared with the same S\&P500 stock market index.
It is also interesting to observe the increasing lag $x(t)$ between the yield rates
and the S\&P500 index from around $0$ month in 2000 to about one year in 2007.
This is followed by a plateau for all yields from FFR to 3Y, that last about 2 years
and is then followed by a decay of the lag thereafter to about half its maximum, i.e., around 6 months.

For the second group of yields with maturities from 5Y to 20Y whose $x(t)$'s are plotted in the right panel of Figs. \ref{Fig:monthly:xt} and \ref{Fig:weekly:xt}, the picture is somewhat different. Before early 2003, the four curves are close to each other with no clear lead-lag structure between them. Then, from 2003 to mid-2007,
a period corresponding to a very bullish upward trend of the stock market boosted
by the favorable low rate of the Fed policy and a booming real-estate bubble,
one can observe that the longer term yields lead clearly the shorter term yields:
\begin{equation}
  x_{\rm{5Y}}(t) < x_{\rm{7Y}}(t) < x_{\rm{10Y}}(t)<x_{\rm{20Y}}(t).
  \label{Eq:TOP:Month:xt:G2a}
\end{equation}
Thereafter, in the reaction to the financial crisis, one observes as for the FFR-3Y yields
that the shorter-term yields lead the long-term yields:
\begin{equation}
  x_{\rm{5Y}}(t) \geq x_{\rm{7Y}}(t) > x_{\rm{10Y}}(t) > x_{\rm{20Y}}(t).
  \label{Eq:TOP:Month:xt:G2b}
\end{equation}
There is much less evidence for a plateau of the lead structure with respect to the S\&P500.

We would also like to mention that a reversal such as the one from (\ref{Eq:TOP:Month:xt:G2a}) to (\ref{Eq:TOP:Month:xt:G2b})
does  not seem to have been documented before.

\section{The S\&P500 leads all yields: evidence from cross-correlation analysis}
\label{S1:Xcorr}

By construction, the traditional cross-correlation analysis \cite{Haugh-1976-JASA} is not adapted
to time varying lead-lag structures. It is however useful to investigate how it performs in the present context
in which the TOP method has diagnosed a significant time varying structure.

\begin{figure}[htb]
 \centering
 \includegraphics[width=5.5cm]{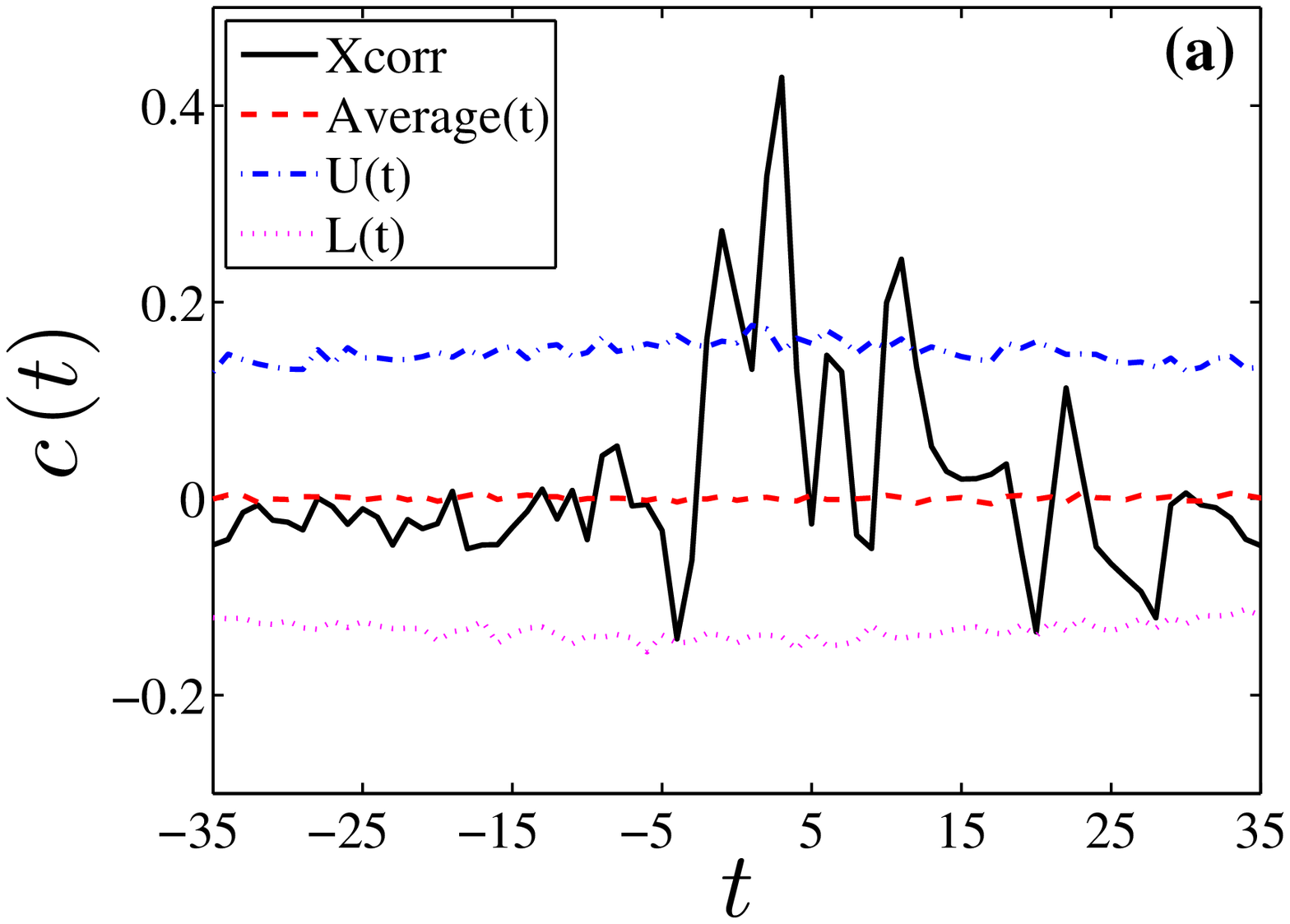}
 \includegraphics[width=5.5cm]{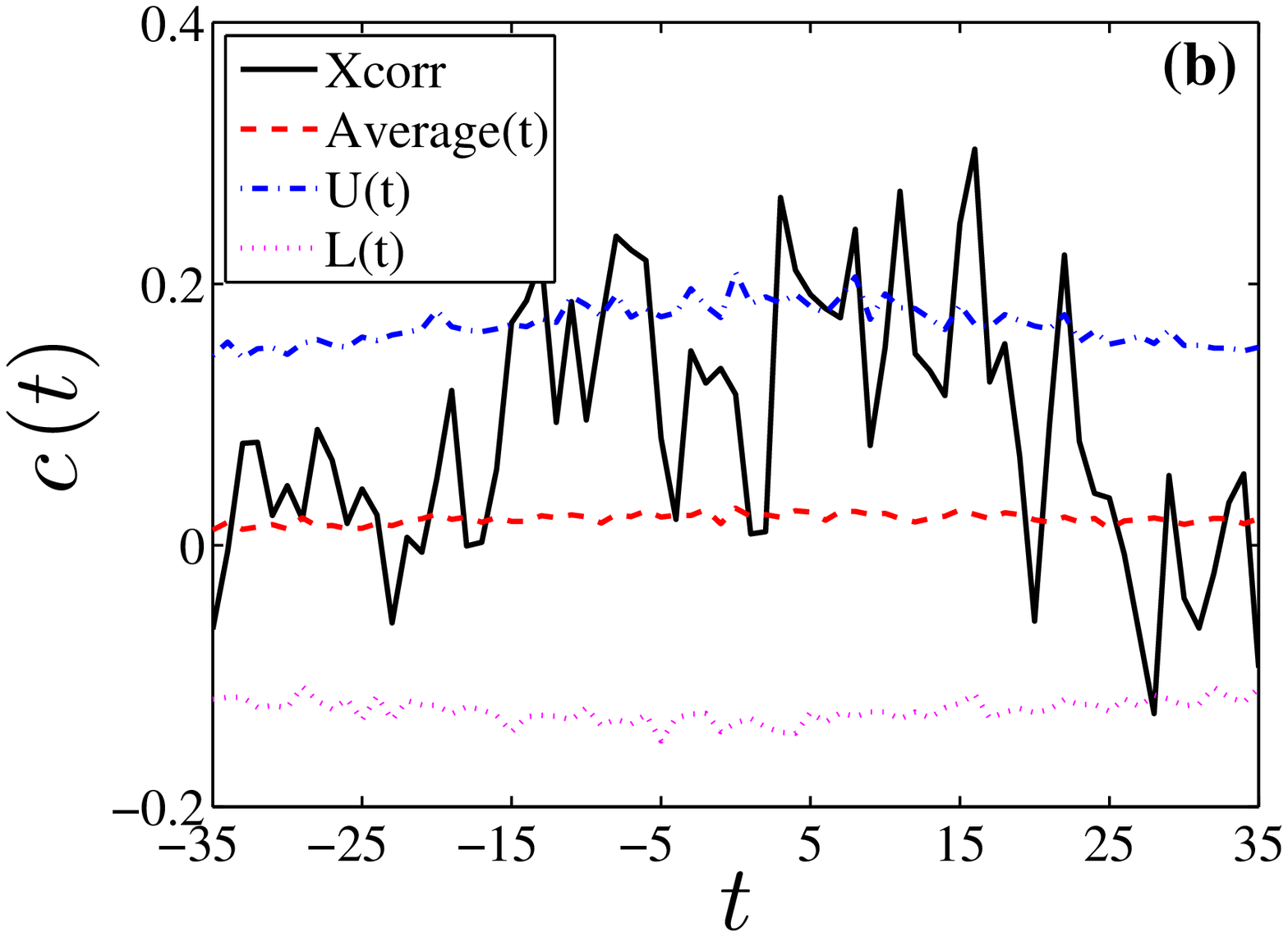}
 \includegraphics[width=5.5cm]{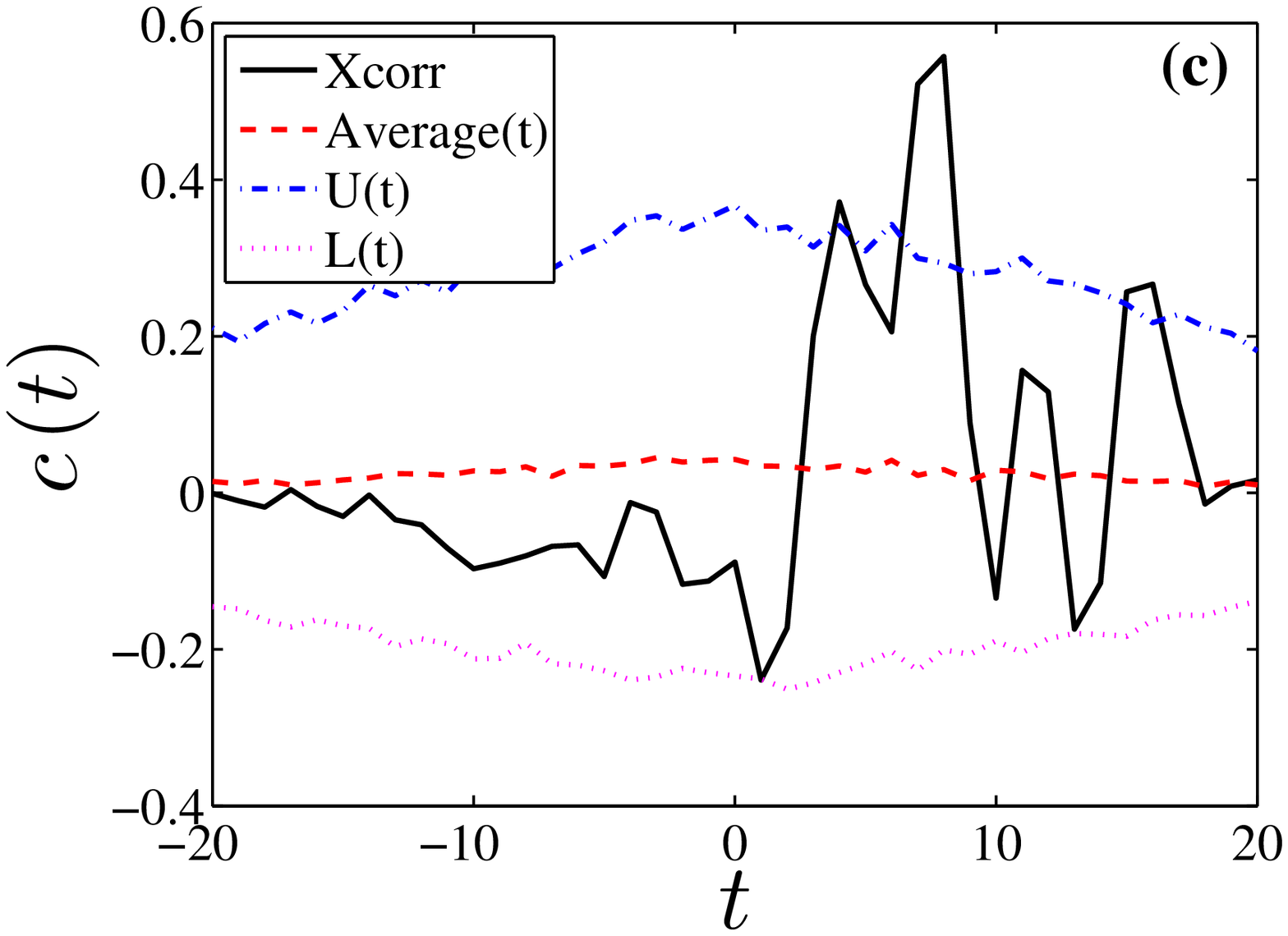}
 \includegraphics[width=5.5cm]{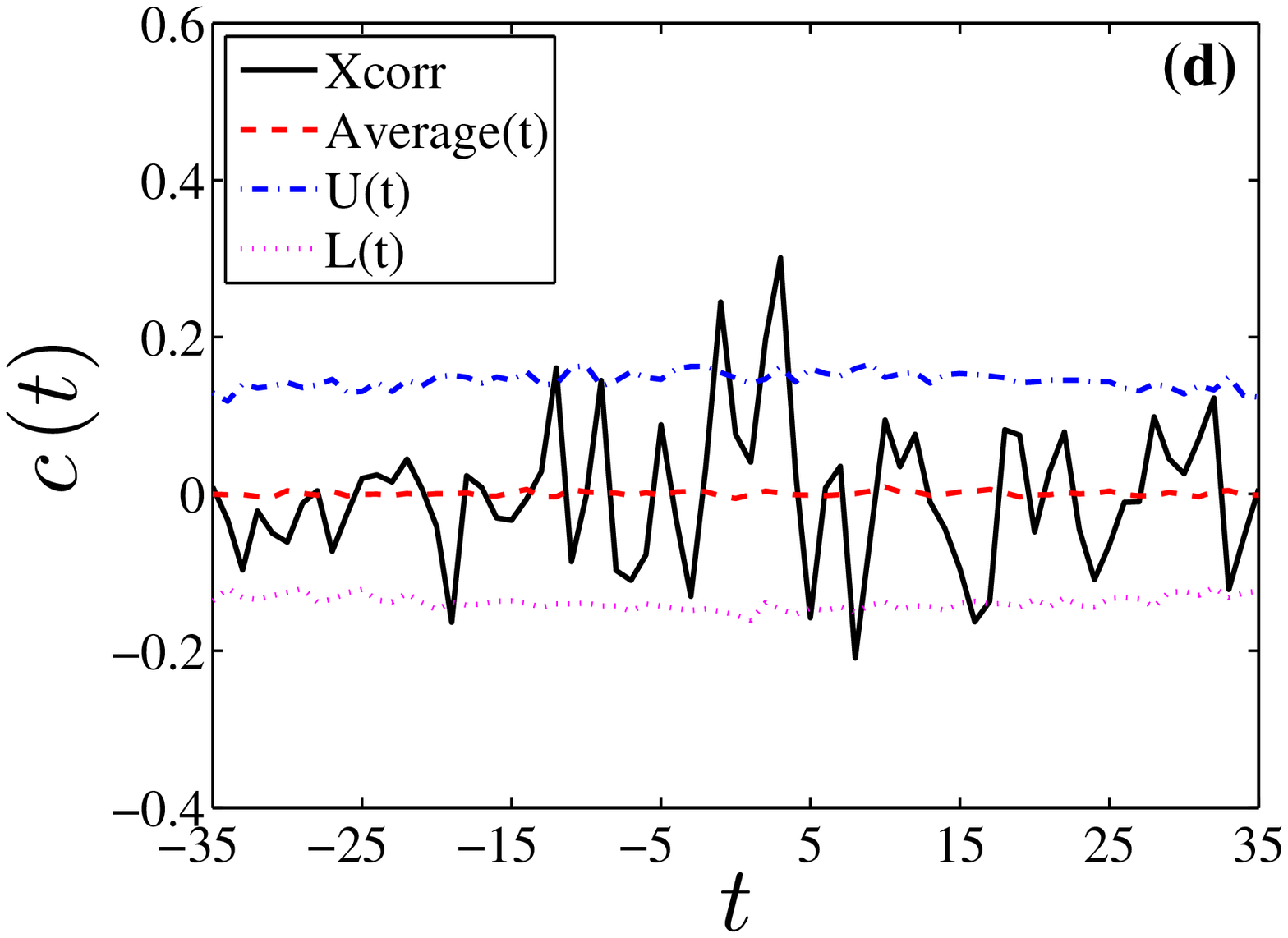}
 \includegraphics[width=5.5cm]{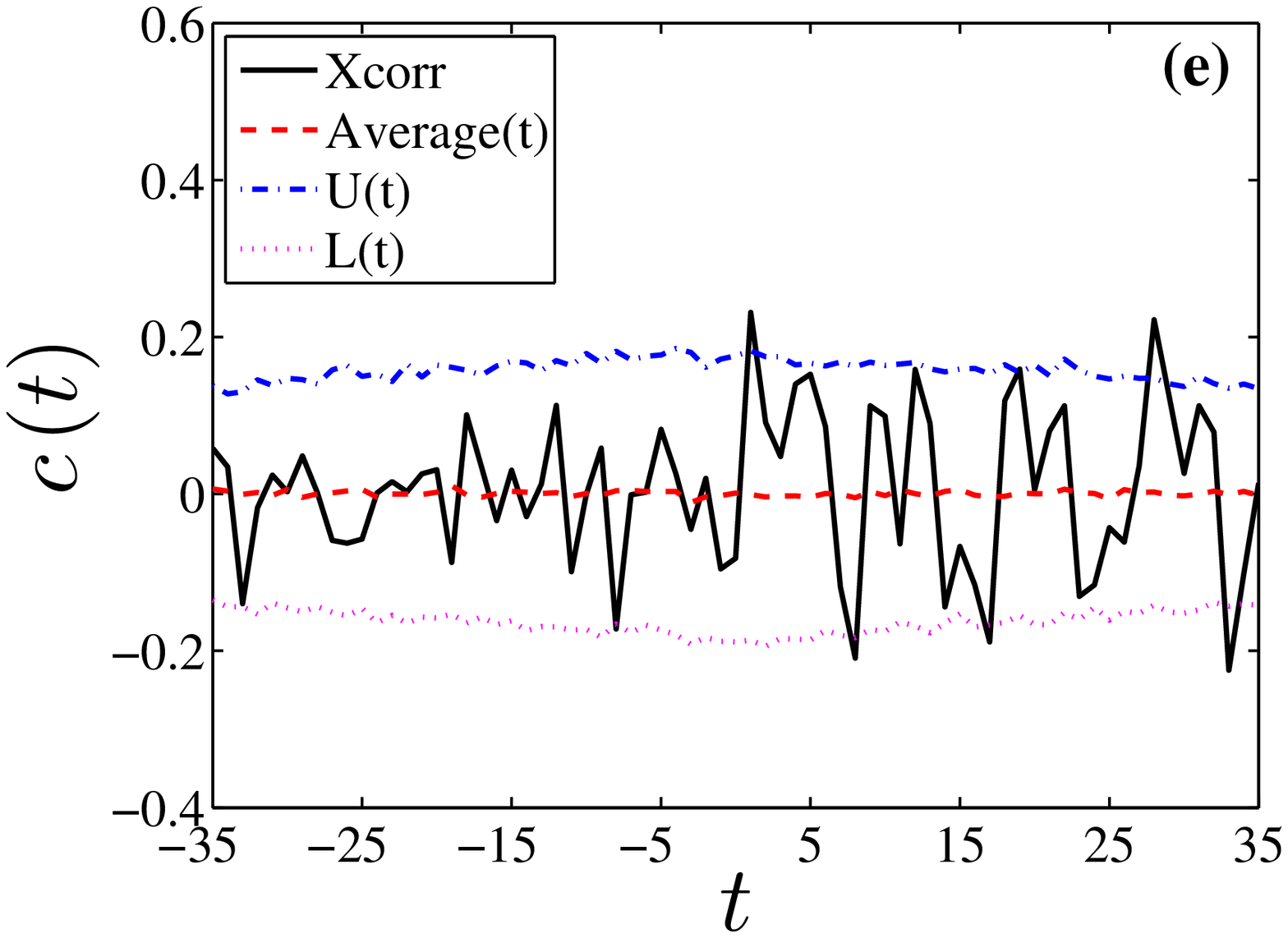}
 \includegraphics[width=5.5cm]{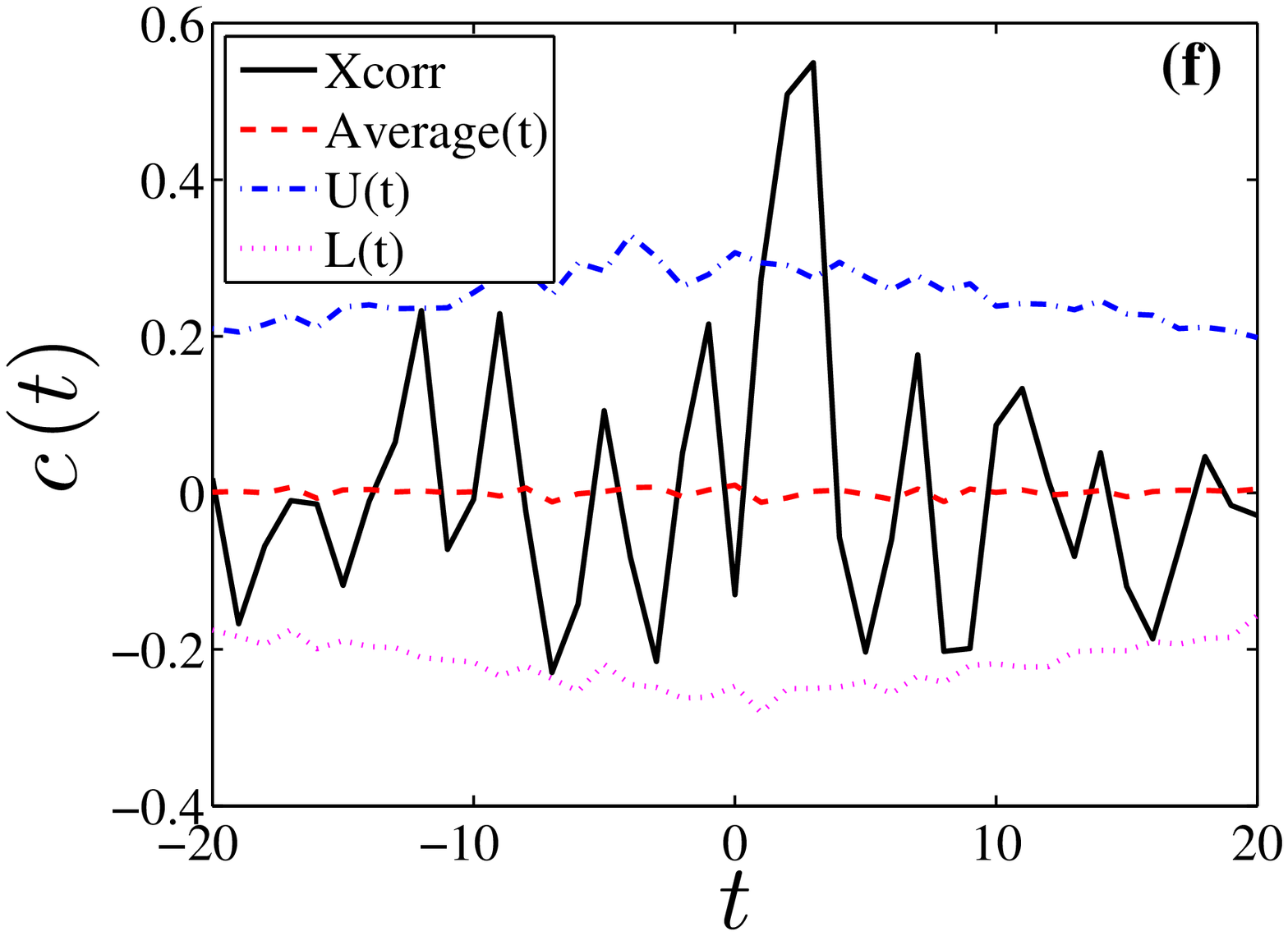}
 \caption{\label{Fig:Xcorr:test} Lagged cross-correlation between the logarithmic return of the S\&P 500 index and the logarithmic difference of the FFR (a-c) and between the logarithmic return of the S\&P 500 index and the logarithmic difference of the 20Y Treasury bond yield (d-f) during different time periods: (a,d) the whole time period from August 2000 to February 2010, (b,e) the time interval from August 2000 to April 2007, and (c,f) the time interval from May 2007 to February 2010. The ordinate axis shows the cross-correlation coefficients $c(t)$. The unit of the lag time $t$ along the abscissa is month.}
\end{figure}

Two representative time series (FFR and 20Y) are presented for illustration. The significance levels of the cross-correlations are evaluated using bootstrapping tests through shuffling the return time series, similar to the analysis for the TOP method described in Section \ref{S1:TOPmethod}B. We use the monthly data in this analysis. For each pair of time series, we analyze the whole time series and two non-overlapping time periods.
The results are shown in Fig.~\ref{Fig:Xcorr:test}. It is obvious that the lagged cross-correlation analysis is not able to characterize the instantaneous evolution of the lead-lag structure evidenced in the previous TOP analysis.  This is not a surprise.

\subsection{(S\&P 500, FFR) pair}

For the (S\&P 500, FFR) pair in the whole time period, the highest peak found in Fig.~\ref{Fig:Xcorr:test}(a), with a positive lag, shows that the FFR lags behind the S\&P 500 index by about 3 months, with a cross-correlation coefficient $c=0.43$, which is significantly positive at the confidence level of 95\%. There are two other peaks that are also significant, one at a negative lag of $t=-2$ month with $c=0.26$ and another at the positive lag $t=8$ month with $c=0.24$. The largest peak with positive lag and highest cross-correlation coefficient $c=0.43$ can be considered as confirming
the main results of the previous section that the stock market changes precede the FFR variations. Due to the fixed lead-lag structure
of the method, the cross-correlation provides only an average coarse representation of the real much richer and dynamical nature
of the lead-lag structure.

For the time period before April 2007, we see many peaks at positive and negative lags $t$ that are significantly different from zero, as shown in Fig.~\ref{Fig:Xcorr:test}(b). It is hard to extract from this plot a clear picture about the lead-lag structure between the S\&P 500 and FFR.
In the presence of large variations of the lead-lag structure, it is not surprising that the cross-correlation analysis is not informative.

For the time period after April 2007, we see in Fig.~\ref{Fig:Xcorr:test}(c) a significant peak at the positive lag $t=8$ month with $c=0.56$. This lag is consistent in magnitude with the average value of the $x(t)$ curve shown in Fig.~\ref{Fig:monthly:xt:test}(a).
This clear signal in the cross-correlation analysis can be explained from the fact that the lead-lag has stabilized
approximately above a value of 6 months,
according to the analysis of the $x(t)$ function shown in Fig.~\ref{Fig:monthly:xt:test}(a) during the time period under investigation.

\subsection{ (S\&P 500, 20Y) pair}

For the (S\&P 500, 20Y) pair in the whole time period, there are two significant peaks around zero lag $t=0$ in Fig.~\ref{Fig:Xcorr:test}(d).

For the time period before April 2007, the signal is ambiguous although we can see several significant peaks in Fig.~\ref{Fig:Xcorr:test}(e).

For the time period after April 2007, we see in Fig.~\ref{Fig:Xcorr:test}(f) only one significant peak at $t=3$ month with $c=0.56$. According to Fig.~\ref{Fig:monthly:xt:test}(b), the lead-lag $x(t)$ decreases from about $6$ to $1$ month. Therefore, these two analyses give consistent results: on average, the S\&P 500 index leads the 20Y Treasury bond yield by about 3 months.

Comparing Fig.~\ref{Fig:Xcorr:test} for the cross-correlation analysis and Fig.~\ref{Fig:monthly:xt:test} for the TOP analysis, we can conclude that the cross-correlation analysis can extract only part of the information and the TOP method is
clearly superior.

\section{Concluding remarks}
\label{S1:Conclusion}

In this work, we have adopted the thermal optimal path method to investigate the dynamics lead-lag structure between the S\&P 500 index of the US stock market and Federal Funds rate, as well as several Treasury bond yields with different maturities. The time period
that has been investigated runs from August 2000 to February 2010.
Both monthly and weekly data have been used and we obtained consistent results. In all cases, the S\&P 500 index is found to lead the FFR and the bond yields. This is quantified by the lead function $x(t)$ found to be positive at a high statistical confidence level determined
by bootstrapping tests. This finding is consistent with and extends significantly a previous work reporting that the US Federal Reserve was ``slaved'' to the stock market during the 2000-2003 US stock market antibubble \cite{Zhou-Sornette-2004b-PA}.

According to the TOP analysis, we observed that the FFR and the Treasury bond yields can be divided into two groups. The first group contains FFR, 3M, 6M, 1Y, 2Y, and 3Y bond yields with short-term maturities and the second group contains 5Y, 7Y, 10Y, and 20Y bond yields with long-term maturities. The lead functions $x(t)$ between the S\&P 500 index and the yields
in each group have very similar quantitative shapes, while they are different at a quantitative level
across the two groups. We found that the short-term yields in the first group lead the long-term yields in the second group before the current financial crisis around 2007 and the inverse relationship holds afterwards, namely
 the long-term yields lead the short-term yields after 2007.

For the first group, the lead function $x(t)$ increases during the time period from 2000 to 2007, followed by a two-year-long plateau, and then plummets in late 2009. We also found that the yields (including FFR) with shorter maturity in the first group have a longer lag behind the
S\&P 500 index than for the longer maturities. In contrast, for the second group, the lead function $x(t)$ increases till 2006 and then decreases. We observed a reversal of the order of the lead functions $x(t)$ among the different maturities in 2007: a yield with shorter maturity has a shorter lag
to the S\&P 500 index before the reversal point and a yield with longer maturity has a shorter lag to the S\&P 500 index after the reversal point. Qualitatively, the reversal phenomenon is coincident with the outbreak of the current financial crisis.

The lag of the FFR to the  S\&P 500 index can be interpreted in the light of comments of
the previous and present Fed chairmen Greenspan and Bernanke that the growth
of stock markets is ``key'' to the recovery and health of the economy.  The evidence
provided here suggests indeed that the FFR policy is in a significant part influenced by the recent
past behavior of the stock markets (stock market $\rightarrow$ Federal Funds rate).
In plain words, the fact that the FFR follows the stock market direction
can be interpreted as a direct attempt to limit its losses and revive it in times of bearish markets
or to stabilize it in times of overly buoyant bubbling markets.

As for the longer maturities, the lag structure with respect to the  S\&P 500 index
reflects (i) a natural link in the term-structure that attach the longer maturities
to the shortest maturity and (ii) the aggregate strategies of investors
facing uncertainties over the long term behavior of the economy \cite{Fatum-Hutchison-1999-JMCB, Brissimis-Magginas-2006-JMonE}.
In the first sub-stage before 04/2007, we
observe the causal relational flow from the stock market $\rightarrow$ Federal Funds rate $\rightarrow$ short-term yields $\rightarrow$ long-term yields, and afterwards, we find the flow from the stock market $\rightarrow$ long-term yields $\rightarrow$ short-term yields $\rightarrow$ Federal funds rate. Thus, the lead-lag structures between the different yields changed after the financial crisis starting in 2007.
This change can be rationalized by the strategies implemented by long-term investors in the face
of growing global market uncertainties, such as central Banks of major Asian countries and pension funds
which are heavily invested in the US long-term Treasury bonds \cite{Warnock-Warnock-2009-JIMF}.
The stern challenges faced by the US economy escalated the uncertainty which cascaded
to exchange rate and inflation. Consequently, the long-term Treasury bonds became quite reactive
to the behavior of stock markets, reflecting the actions of these long-term investors ``flying to safety'':
a plunge in the stock markets led to strong demand for the supposedly safe US Treasury bonds, pushing down mechanically
the corresponding yields. This suggests that the long-term investors have been more reactive
and mindful of the signals provided by the financial stock markets than the Federal Reserve itself after the start of the
financial crisis.
This may be due to the more complex agenda as well as the delicate role of the Federal Reserve, which
has to take into account the impact of its interventions \cite{Baeriswyl-Cornand-2010-JMonE}.
Caution and prudence on the part of the Fed in a time of high uncertainty
may thus be the reason for this inversion of the lead-lag relationship between changes of yields of different maturities.
However, the robust lead of the S\&P 500 stock market index with respect to yields
of all maturities remains the most important stylized fact unearthed by our study.

\begin{acknowledgments}
KG and SWC acknowledges financial support from the Chinese Academy of Sciences Foundation for Planning and Strategy Research (Grant No. 0929015ED2). WXZ acknowledges financial support from the ``Shu Guang'' project (Grant No. 2008SG29) sponsored by Shanghai Municipal Education Commission and Shanghai Education Development Foundation, the National Natural Science Foundation of China (Grant No. 11075054), and the Fundamental Research Funds for the Central Universities. DS acknowledges financial support from the ETH Competence Center ``Coping with crises in complex socio-economic systems'' (CCSS) through ETH Research Grant CH1-01-08-2 and ETH Zurich Foundation and ETH Research Grant ETH-31 10-3 ``Testing the predictability of financial bubbles and of systemic instabilities''.

\end{acknowledgments}

%\bibliography{Bibliography}
\bibliography{E:/Papers/Auxiliary/Bibliography}

\end{document}